\documentclass[pra,aps,twocolumn,superscriptaddress,english,longbibliography]{revtex4-2}

\usepackage{bm,amssymb,amsmath,mathrsfs,amstext,latexsym,physics}
\usepackage{cancel,hhline}
\usepackage[usenames,dvipsnames]{xcolor}
\usepackage[colorlinks=true,citecolor=blue,urlcolor=blue,linkcolor=blue]{hyperref}

\usepackage[caption=false,subrefformat=parens,labelformat=empty]{subfig}
\usepackage{graphicx,tikz}

\newcommand{\der}[2]{\frac{\mathrm{d} #1}{\mathrm{d} #2}}
\newcommand{\pd}[2]{\frac{\partial #1}{\partial #2}}
\newcommand{\fd}[2]{\frac{\delta #1}{\delta #2}}

\renewcommand{\tilde}[1]{\widetilde{#1}}

\newcommand{\figpanel}[2]{\hyperref[#1]{\ref{#1}#2}}

\begin{document}

\title{Non-Adiabatic Quantum Optimization for Crossing Quantum Phase Transitions}

\author{Andr\'as Grabarits\href{https://orcid.org/0000-0002-0633-7195}{\includegraphics[scale=0.05]{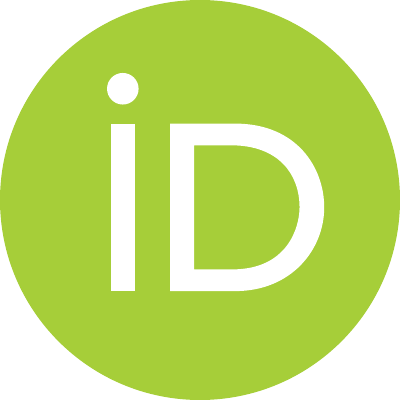}}}
\email{andras.grabarits@uni.lu}
\affiliation{Department  of  Physics  and  Materials  Science,  University  of  Luxembourg,  L-1511  Luxembourg, Luxembourg}

\author{Federico Balducci\href{https://orcid.org/0000-0002-4798-6386}{\includegraphics[scale=0.05]{orcidid.pdf}}}
\email{fbalducci@pks.mpg.de}
\affiliation{Department  of  Physics  and  Materials  Science,  University  of  Luxembourg,  L-1511  Luxembourg, Luxembourg}
\affiliation{Max Planck Institute for the Physics of Complex Systems, N\"othnitzer Str. 38, 01187 Dresden, Germany}
\author{Barry C. Sanders\href{https://orcid.org/0000-0002-8326-8912}{\includegraphics[scale=0.05]{orcidid.pdf}}}
\affiliation{Institute for Quantum Science and Technology, University of Calgary, Alberta T2N 1N4, Canada}
\author{Adolfo del Campo\href{https://orcid.org/0000-0003-2219-2851}{\includegraphics[scale=0.05]{orcidid.pdf}}}
\affiliation{Department  of  Physics  and  Materials  Science,  University  of  Luxembourg,  L-1511  Luxembourg, Luxembourg}
\affiliation{Donostia International Physics Center,  E-20018 San Sebasti\'an, Spain}

\date{\today}


\begin{abstract}
    We consider the optimal driving of the ground state of a many-body quantum system across a quantum phase transition in finite time. In this context, excitations caused by the breakdown of adiabaticity can be minimized by adjusting the schedule of the control parameter that drives the transition. Drawing inspiration from the Kibble-Zurek mechanism, we characterize the timescale of onset of adiabaticity for several optimal control procedures. Our analysis reveals that schedules relying on local adiabaticity, such as Roland-Cerf's local adiabatic driving and the quantum adiabatic brachistochrone, fail to provide a significant speedup over the adiabatic evolution in the transverse-field Ising and long-range Kitaev models. As an alternative, we introduce a novel framework, Non-Adiabatic Quantum Optimization (NAQO), that, by exploiting the Landau-Zener formula and taking into account the role of higher-excited states, outperforms schedules obtained via both local adiabaticity and state-of-the-art numerical optimization. NAQO is not restricted to exactly solvable models, and we further confirm its superior performance in a disordered non-integrable model.
\end{abstract}

\maketitle

\section{Introduction}
\label{sec:intro}

Quantum phase transitions (QPT) play a dominant role in the understanding of quantum matter in and out of equilibrium~\cite{Sachdev2011Quantum,Sachdev2023Quantum}. By definition, at a QPT the ground-state properties change abruptly as the external parameters are varied. 
In addition, when the transition is of second order, the energy gap separating the ground state from the first excited state closes. As a result, the coherent driving across a second-order QPT is non-adiabatic and results in excitations above the ground state~\cite{Dziarmaga10,Polkovnikov2011Colloquium,DZ14}. The universality class of the QPT determines the number density of such excitations, according to the celebrated Kibble-Zurek (KZ) prediction $n \sim T^{-\alpha_\mathrm{KZ}}$ ~\cite{Kibble76a,Kibble76b,Zurek85,Zurek96c}, where $T$ is the time scale in which the transition is driven and $\alpha_\mathrm{KZ} = d\nu/(1+z\nu)$, with $d$ the dimensionality of the system, $\nu$ the correlation length critical exponent and $z$ the dynamic critical exponent of the QPT~\cite{Polkovnikov2005Universal,Damski05,Dziarmaga2005Dynamics,Zurek2005Dynamics}.

Reducing excitation formation across a QPT is desirable for a variety of applications in quantum technologies. Examples include the preparation of novel phases of matter (e.g.,\ in quantum simulators) and quantum annealing, where the KZ mechanism provides a heuristic guideline for simple energy landscapes~\cite{Gardas18,Keesling2019,Weinberg20,Bando20,King22,King2024sup,Andersen2024}. Small values of $\alpha_\mathrm{KZ}$ require exceedingly long values of $T$ to approach the adiabatic limit. This has led to a variety of proposals to circumvent the KZ scaling.

One approach relies on the use of auxiliary controls, e.g.,\ based on counterdiabatic driving~\cite{Demirplak2003Adiabatic,Demirplak2005Assisted,Demirplak2008Consistency,Berry2009Transitionless}. The auxiliary controls required to suppress a QPT are generally nonlocal in real space and involve multiple-body interactions, which hinders their practical implementation~\cite{delcampo12,Takahashi2013,Saberi14,Damski2014}. Efforts to approximate such controls by local few-body terms have been pursued by a variety of approaches~\cite{Saberi14,Sels2017Minimizing,Claeys2019Floquet,Zhang2024analogCD}. 
Digital quantum simulation based on the gate model facilitates the implementation of the required controls and has led to the development of digitized counterdiabatic quantum algorithms for quantum optimization ~\cite{Hegade2021,YaoLinBukov21,Chandarana2022Digitized,Wurtz2022Counterdiabaticity,McKeever2024Towards}. Despite limitations resulting from  Trotter errors and gate fidelities, theoretical proposals have been accompanied by their experimental demonstration \cite{Chandarana2023protein,Chandarana2023photonicCD,Li2024jpmorgan,Cadavid2024}. An alternative to counterdiabatic driving relies on making the critical point spatially dependent: both in classical~\cite{KV97,Zurek09,delcampo10,delCampo2011} and quantum systems~\cite{DM10,DM10b,Francuz16,GomezRuiz19,sokolov2024inhomogeneous}, this can help mitigate the formation of excitations~\cite{DKZ13}, as demonstrated in various experiments \cite{Ulm13,Pyka13,Yi2020Exploring,Kim2022}.

In the absence of auxiliary or spatially local controls, previous studies have considered the modulation in time of the control parameter $g(t)$ by non-linear schedules~\cite{Diptiman08}. Specifically, for a power-law dependence $g(t)=(t/T)^r$, the power-law KZ scaling is modified to $n \sim T^{-dr\nu/(1+dr\nu)}$: by optimizing the value of $r$, the excitation density is improved with respect to the standard KZ scaling, and gets close to $n \sim T^{-1/z}$~\cite{Barankov08,Power13,Wu2015Optimal}. Boundary cancellation techniques provide an alternative approach to tailor the schedule $g(t)$: provided that the first $k$ derivatives of $g(t)$ vanish at the beginning and end of the process, nonadiabatic excitations are bounded by $C_k/T^{k+1}$ for some positive constant $C_k$~ \cite{Garrido62,Nenciu1993,Hagedorn02,Lidar2009Adiabatic,Wiebe2012,Ge16}. Beyond such analytical insights, numerical approaches for optimal control have been proposed to design optimal driving schedules~\cite{Khaneja2005Optimal,Caneva09,Doria11,Caneva11,Boscain2021Introduction,DAlessandro2021Introduction}. 

In this work, we exploit the KZ mechanism to characterize the performance of optimal control methods applied to schedules $g(t)$ that drive the QPT in finite time. We show how methods that rely only on the adiabatic condition for the ground state, such as local adiabatic driving (LAD) by Roland and Cerf~\cite{Roland2002Quantum} and the closely related quantum adiabatic brachistochrone (QAB)~\cite{Rezakhani2009Quantum, Campbel_MinAction2022}, fail to deliver schedules $g(t)$ that perform well at moderate driving rates, and thus cannot be used to speed up significantly the quantum evolution. Our understanding is based on a semi-analytical treatment in the transverse-field Ising model (TFIM) and long-range Kitaev model (LRKM), which, thanks to their free-fermionic nature, allow for the diagonalization of the excitations in well-defined momentum sectors. We then try to remedy the shortcomings of the QAB by generalizing its structure to each momentum mode: our extended quantum adiabatic brachistochrone (eQAB), however, does not provide optimal schedules $g(t)$ that show significant speed-up. A different approach, based on the Landau-Zener (LZ) formula rather than the adiabatic condition, is needed to obtain a significant improvement, as we show in the TFIM and even more significantly in the LRKM. To this end, we introduce the framework of Non-Adiabatic Quantum Optimization (NAQO) that takes into account excited states in combination with the LZ formula. Such an approach outperforms the optimized non-linear passage (ONLP) ~\cite{Barankov08} and a purely numerical optimization method, the Chopped Random Basis (CRAB) ansatz~\cite{Doria11,Caneva11,Muller2022One}. The method is benchmarked  as well on a non-integrable, disordered spin model, showing promising results for the accessible system sizes.

Acronyms used throughout the work are listed in Table. \ref{ListAcronyms}. The paper is organized as follows. In Sec.~\ref{sec:TFIM}, we define the problem under study: we introduce the TFIM, diagonalize it in momentum space (Sec.~\ref{sec:TFIM_diagonalization}) and review the basics of the KZ  mechanism (Sec.~\ref{sec:TFIM_KZ}). In Sec.~\ref{sec:opt_contr_adiab}, we review how the adiabatic condition can be used to tailor locally adiabatic schedules; specifically, we consider LAD (Sec.~\ref{sec:LAD}) and the QAB (Sec.~\ref{sec:QAB}). In Sec.~\ref{sec:LAD=QAB}, we show an interesting connection between the two approaches. In Sec.~\ref{sec:opt_contr_beyond_adiab}, we apply optimization procedures to the TFIM that go beyond the adiabatic condition, namely the eQAB (Sec.~\ref{sec:eQAB}),  NAQO (Sec.~\ref{sec:NAQO}) and ONLP  (Sec.~\ref{sec:ONLP}). In Sec.~\ref{sec:CRAB}, we benchmark our NAQO approach against fully numerical, quantum optimal control strategies. In Sec.~\ref{sec:LRK}, we perform a similar analysis in the LRKM. We introduce the model in Sec.~\ref{sec:LRK_diagonalization} and study the crossing of its quantum critical point both in the KZ universality regime (Sec.~\ref{sec:LRK_KZregime}) and in the dynamical universality regime (Sec.~\ref{sec:LRK_dyn_univ_regime}). We draw our conclusions in Sec.~\ref{sec:conclusions}.


\begin{table}[t]
    \begin{tabular}{ll}
        CRAB & Chopped Random Basis \\
        KZ & Kibble-Zurek\\
        LAD & Local Adiabatic Driving\\
        LRKM & Long-Range Kitaev Model\\
        LZ & Landau-Zener\\
        NAQO \hspace{0.2cm} & Non-Adiabatic Quantum Optimization\\
        MZM & Majorana Zero Mode\\
        TFIM & Transverse-Field Ising Model\\
        TLS & Two-Level System\\
        ONLP & Optimal Non-Linear Passage\\
        QAB & Quantum Adiabatic Brachistochrone \\ 
        eQAB &  Extended Quantum Adiabatic Brachistochrone \\
        QPT & Quantum Phase Transition \\
    \end{tabular}
    \label{ListAcronyms}
    \caption{List of acronyms used in the text. 
    }
\end{table}

\section{Transverse-Field Ising Model}
\label{sec:TFIM}


\subsection{Definition}
\label{sec:TFIM_def}

The TFIM describes a  spin chain with  Hamiltonian~\cite{Suzuki2012Quantum}
\begin{equation}
	\label{eq:H_TFIM}
	\hat{H}(t) = - J \sum_{j=1}^L \left[ \hat{\sigma}^z_j \hat{\sigma}^z_{j+1} + g(t) \hat{\sigma}_j^x \right],
\end{equation}
where $\hat{\sigma}_j^{x,y,z}$ are Pauli matrix operators acting on site $j$. The coupling $J > 0$ favors ferromagnetic alignment, and it is set $J\equiv 1$ to fix the energy scale; $g(t)$ plays the role of an effective dimensionless magnetic field. The values $g_c = \pm 1$ are critical points separating the ferromagnetic phase, in which spins align as a result of the strong spin-spin interactions ($|g| < 1$), from the phases with paramagnetic order ($|g| > 1$), where the external magnetic field determines the spin configuration. Due to its simplicity, the TFIM is regarded as the minimal example to describe equilibrium QPTs with discrete symmetry breaking~\cite{Sachdev2011Quantum}, as well as the crossing of QPTs in a nonequilibrium context~\cite{Zurek2005Dynamics,Dziarmaga2005Dynamics,delcampo12,delCampo2018Universal,Bando20,King22,Balducci2023Large}.

In this work, we are interested in driving the ground state of the TFIM from its ferromagnetic phase to the paramagnetic one. To this end, consider a schedule $g(t)$ with $t \in [-T/2,T/2]$, where the time $T$ sets the total duration of the process. It is natural to consider schedules satisfying
\begin{equation}
    g(-T/2) = g_0 > g_c, \qquad
    g(T/2) = 0;
\end{equation}
e.g., with $g_0=2$. 
The goal is to determine the optimal functional form $g(t)$ 
to minimize excitation formation upon crossing the QPT. 

As usual in optimization problems, the first step is to define the cost function. There are multiple choices at hand for the problem under consideration. For instance, one can choose as a cost function the mean energy upon completion of the schedule
\begin{equation}
    E = \bra{\Psi(T/2)} \hat{H}(T/2) \ket{\Psi(T/2)},
\end{equation}
where $\ket{\Psi(T/2)}$ is obtained by integrating the time-dependent Schr\"odinger equation
\begin{equation}
    \begin{cases}
        i \partial_t \ket{\Psi(t)} = \hat{H}(t) \ket{\Psi(t)} \\
        \ket{\Psi(-T/2)} = \ket{E_0(-T/2)}.
    \end{cases}
\end{equation}
Above, $\ket{E_n(t)}$ stands for the $n$-th eigenstate of $\hat{H}(t)$, so that $\ket{E_0(-T/2)}$ is the initial ground state.

Alternatively, one can consider the density of topological defects at the end of the schedule. Due to the $\mathbb{Z}_2$ symmetry of the model, topological defects correspond to kinks. The defect number operator reads 
\begin{equation}
    \hat{N} = \frac{1}{2} \sum_{j=1}^L  \left(1-\hat{\sigma}^z_j \hat{\sigma}^z_{j+1} \right),
\end{equation}
so that the final value of the defect density is given by
\begin{equation}
    n = \frac{1}{L} \bra{\Psi(T/2)} \hat{N} \ket{\Psi(T/2)}.
\end{equation}
Note that for $g(T/2) = 0$, the Hamiltonian contains solely ferromagnetic interactions and commutes with the defect number operator: $[\hat{H}(T/2), \hat{N}] = 0$. Therefore, the mean energy and the defect density are related to each other by $E =  (1-2Ln)$. 

We do not consider in this work the possibility of using the fidelity $F = |\braket{E_0(T/2)}{\Psi(T/2)}|^2$ (or its logarithm) as a cost function since it is experimentally much more difficult to measure in a many-body setting. 

\begin{figure}
    \includegraphics[width=\columnwidth]{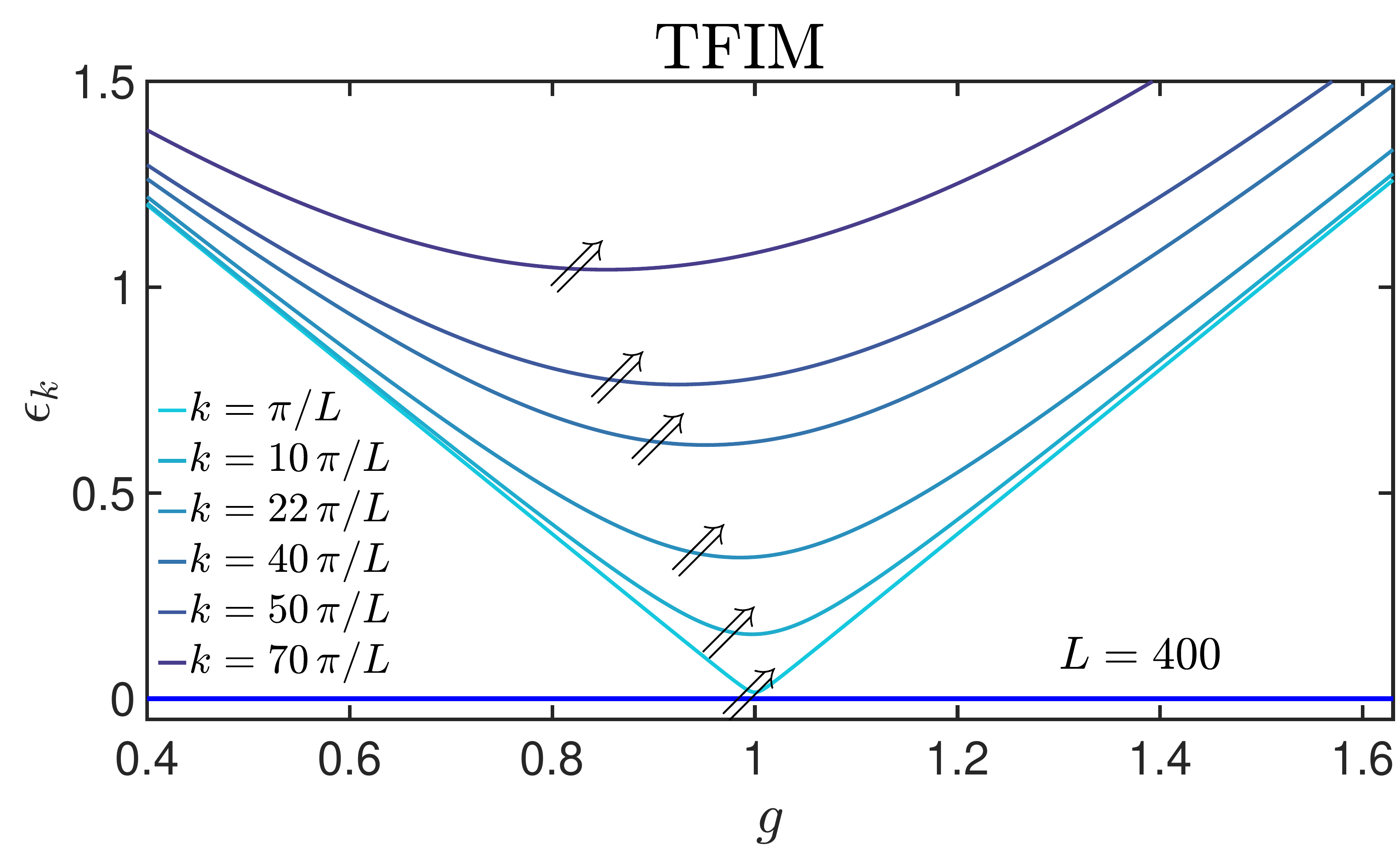}
    \caption{Schematic plot of the energy gaps in each momentum mode of the TFIM. The avoided crossings (signaled by a double arrow) take place under the umbrella of the zero mode and thus remain unaccounted by the local adiabatic condition, see Sec.~\ref{sec:opt_contr_adiab}. By contrast, the NAQO approach of Sec.~\ref{sec:NAQO} tames each of the avoided crossings individually.}
    \label{fig:TFIM_schematic}
\end{figure}

\subsection{Diagonalization and avoided level crossings}
\label{sec:TFIM_diagonalization}

Upon using a Jordan-Wigner transformation, the Hamiltonian~\eqref{eq:H_TFIM} is reduced to that of a set of independent two-level systems (TLSs), one for each (pseudo-) momentum sector~\cite{Suzuki2012Quantum},
\begin{align}
	\label{eq:H_TFIM_momentum}
	\hat{H} &= 2 \sum_{k>0} \hat{\psi}_k^\dagger \left[ (g(t)-\cos k)\tau^z + \sin k \, \tau^x \right] \hat{\psi}_k \\
    &= 2 \sum_{k>0} \hat{\psi}_k^\dagger H_k(t) \hat{\psi}_k,
\end{align}
where $\hat{\psi}_k := (\hat{c}_k, \hat{c}_{-k}^\dagger)^T$ is a vector of creation and annihilation operators for fermions of momentum $k$, and $\tau^{x,y,z}$ are another set of Pauli matrices, acting on the internal space of $\hat{\psi}_k$. By diagonalizing the 2$\times$2 matrices $H_k(t)$, one finds the single-particle eigenenergies
\begin{equation}
    \label{eq:TFIM_e_k}
    \epsilon_{k}(t) = \pm 2 \sqrt{1+g^2(t) -2g(t)\cos k};
\end{equation}
see also Fig.~\ref{fig:TFIM_schematic}. The eigenenergy of the system is recovered upon adding up all the contributions of different momenta: $E_n(t) = \sum_{k>0} s_{n,k} \epsilon_k(t)$ with $s_{n,k} = \pm 1$ defining the occupation of the modes. 

For momenta $k<\pi/2$, the single-particle levels $\epsilon_k(t)$ undergo avoided level crossings: their position $t_c(k)$ is identified by
\begin{equation}
    \label{eq:AvCross_pos}
    \pd{\epsilon_k}{g} = 0 \implies g_c = \cos k
    \implies t_c(k) = \arccos k;
\end{equation}
see Fig.~\ref{fig:TFIM_schematic}. Notice that, for finite system size, the minimum of the gap is not at $g=1$, but at $g=\cos k_0$, with $k_0=\pi/L$.

\subsection{Kibble-Zurek mechanism}
\label{sec:TFIM_KZ}

As for the energy, also the average number of defects simplifies when expressed in the momentum basis \cite{Dziarmaga2005Dynamics}. In particular, 
\begin{equation}
    n = \frac{1}{L} \sum_{k>0} p_k,
\end{equation}
where $p_k$ is the excitation probability in the sector $k$. Assuming that the driving is a linear ramp $g(t) = g_0(1-t/T)$, such probabilities can be computed approximately in various regimes. First, in the adiabatic limit $T \gg [E_1-E_0]^{-2} \sim L^2$, it holds that $p_k \approx 0$ for all $k$, and thus $n \approx 0$: this is the consequence of the adiabatic theorem. For faster ramps, the LZ formula holds~\cite{Damski05,Dziarmaga2005Dynamics,DamskiZurek06}, 
\begin{equation}
    p_k \approx e^{-2\pi k^2 T},
\end{equation}
from which
\begin{equation}
    n \approx \frac{1}{2\pi} \int_0^{\pi/2} \mathrm dk \, p_k \sim T^{-1/2}.
\end{equation}
This result is in agreement with the original argument of the KZ mechanism, predicting $n \sim T^{-d\nu/(1+z\nu)}$ in $d$ dimensions for point-like defects. Here, the critical exponents read $\nu=z=1$, resulting in the KZ exponent $d\nu / (1+z\nu) = 1/2$. At fast quenches, there is a universal breakdown of the KZ scaling~\cite{Zeng2023Universal}. In the limit of sudden quenches, $p_k = \cos^2(k/2) + \mathcal{O}(T^{1/2})$ for all $k$, leading to a saturation $n \approx 1/4$.

\section{Optimal control based on the adiabatic condition}
\label{sec:opt_contr_adiab}

In a seminal work, Roland and Cerf~\cite{Roland2002Quantum} proposed a simple recipe to realize adiabatic schedules in a time shorter than the one set by a naive application of the adiabatic condition. Instead of requiring the total duration of a linear schedule to be longer than the adiabatic time, 
\begin{equation}
    \label{eq:T_ad}
    T > T_\mathrm{ad} = \frac{\max_t |\ev{\partial_t \hat{H}}{\Psi(t)}|}{\min_t [E_1(t) - E_0(t)]^2},
\end{equation}
they argued that it is sufficient to require ``local adiabatic driving'' (LAD), encoded in the condition
\begin{equation}
    \label{eq:LAD}
    \frac{|\mel{E_0(t)}{\partial_t \hat{H}}{E_1(t)}|}{[E_1(t) - E_0(t)]^2} < \varepsilon,
\end{equation}
with $\varepsilon$ being a small adjustable parameter. This way, the instantaneous rate of change of the external parameter is kept slow enough relative to the instantaneous gap, and the schedule can speed up when the gap is large.

Another approach, based on the same intuition, is provided by the quantum adiabatic brachistochrone (QAB)~\cite{Rezakhani2009Quantum,Campbel_MinAction2022}. Instead of requiring that Eq.~\eqref{eq:LAD} holds at each time, one can define a cost function (action) for the whole schedule as~\footnote{Some authors use in the argument of the integral a different power, e.g.,\ $S = \int \mathrm{d}t \, \| \partial_t H \| / [E_1(t) - E_0(t)]^2$.}
\begin{equation}
    \label{eq:QAB}
    S_\mathrm{QAB} = \int \mathrm{d}t \, \frac{\| \partial_t \hat{H} \|^2}{[E_1(t) - E_0(t)]^4},
\end{equation}
where $\| \bullet \|$ is the Hilbert-Schmidt norm. The optimal driving $g_\mathrm{QAB}(t)$ can then be found by minimizing the action.

In the next sections, we inspect the performance of the two approaches outlined above in the TFIM: the LAD is treated in Sec.~\ref{sec:LAD} and the QAB in Sec.~\ref{sec:QAB}. Taking inspiration from the KZ mechanism, we look at the scaling of the defect number with the final time $T$ when the schedule $g(t)$ is not a linear function but instead determined by Eqs.~\eqref{eq:LAD} or \eqref{eq:QAB}. We also compute the timescale of the onset of adiabaticity, i.e., \ the time $T_\mathrm{ons}$, after which the defect number becomes of $\mathcal{O}(1)$. This timescale is not to be confused with $T_\mathrm{ad}$, which always refers to the global adiabatic condition, Eq.~\eqref{eq:T_ad}. 

In short, both approaches fail to suppress excitations efficiently as they are governed by the adiabatic condition of the zero mode. From a broader perspective, this is the consequence of the underlying many-body structure, allowing for a large variety of pathways to excite the system.
In Sec.~\ref{sec:LAD=QAB}, we prove a simple connection between the two schemes, LAD and QAB, that shows how the two are interchangeable under some approximation.

\subsection{Local adiabatic driving (LAD)}
\label{sec:LAD}

Within the LAD approach, we note that Eq.~\eqref{eq:LAD} can be rewritten as
\begin{equation}
    \label{eq:LAD_gdot}
    |\dot g(t)| < \varepsilon \frac{[E_1(t)-E_0(t)]^2}{\big| \mel{E_0(t)}{\partial_g \hat{H}(t)}{E_1(t)} \big|}.
\end{equation}
Using the exact solution of the TFIM (Sec.~\ref{sec:TFIM_diagonalization}), it is possible to determine the explicit expressions of the gap and the matrix element to find
\begin{equation}
    \label{eq:LAD_TFIM}
    |\dot{g}| < \frac{8\varepsilon}{\sin k_0} (1+g^2-2g\cos k_0)^{3/2},
\end{equation}
where the absolute value is needed because the schedule from the paramagnet to the ferromagnet makes $g$ monotonically decrease in time. Equation \eqref{eq:LAD_TFIM} upper bounds the local speed of the schedule. Notice that both the gap and the matrix element were determined solely by the zero-mode TLS with $k_0=\pi/L$. Replacing the inequality above with an equality, the differential equation for $g$ can be solved by separation of variables:
\begin{equation}
    \label{eq:LAD_TFIM_integrated}    
    -\varepsilon' (t-t_0) = \frac{g-\cos k_0}{\sqrt{1+g^2-2g\cos k_0}},
\end{equation}
where $\varepsilon' = 8 \varepsilon \sin k_0$ and $t_0$ is the integration constant. A minus sign was added upon removal of the absolute value, taking into account that $\dot g<0$. At this point, the two free parameters $\varepsilon'$ and $t_0$ can be fixed by imposing $g(-T/2) = g_0 = 2$, $g(T/2)=0$. While their exact expression is rather cumbersome, at large $L$ it holds 
\begin{equation}
    \label{eq:epsilon_t0_LAD}
    \varepsilon'= \frac{1}{T} \left[ 2 + \mathcal{O}\left( \frac{1}{L^2} \right) \right], \quad
    t_0 = T \left[ \frac{\pi^4}{4L^4} + \mathcal{O}\left( \frac{1}{L^6} \right) \right].
\end{equation}
To leading order, the schedule becomes symmetric around  $t=0$, and the small parameter $\varepsilon$ sets the inverse of the annealing time. Solving Eq.~\eqref{eq:LAD_TFIM_integrated} for $g$, one finds
\begin{equation}
    \label{eq:g_LAD}
    g_\mathrm{LAD}(t) = \cos k_0 - \frac{\varepsilon' (t-t_0)\sin k_0}{\sqrt{1-\varepsilon'^2 (t-t_0)^2}}.
\end{equation}
This is the functional form of the optimal annealing schedule according to the LAD scheme.

\begin{figure}
    \includegraphics[width=\columnwidth]{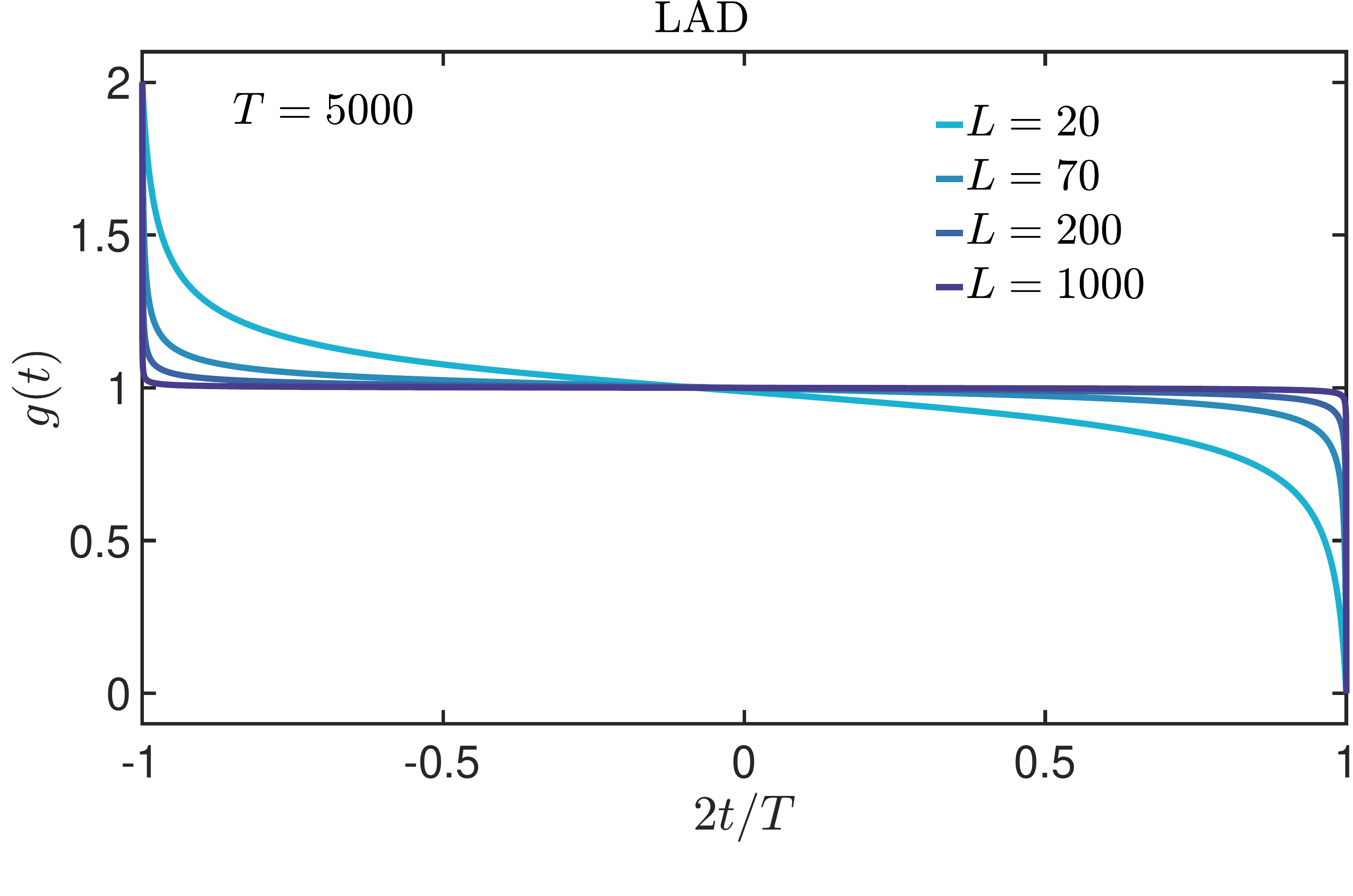}
    \caption{Schedule $g(t)$ for the optimal crossing of the QPT in the TFIM, according to the LAD. One can see that the curves become flatter and flatter as the system size $L$ increases, with sharp jumps at the endpoints: this leads to significant excitation formation, as explained in the main text, Sec.~\ref{sec:LAD}.}
    \label{fig:LADprotocol}
\end{figure}
The critical point is traversed at $t = 0 + \mathcal{O}(1/L^4)$, with a slope $\dot{g}_\mathrm{LAD} \approx - 2\pi/LT$. The effect of LAD is thus to slow down near the critical point while going very fast away from it, in order to arrive in finite time from $g_0$ to 0.
The shape of the $g_\mathrm{LAD}(t)$ acquires a dependence on $L$, as shown in Fig.~\ref{fig:LADprotocol}. Even though to leading order the schedule depends only on $t/T$, as follows from the expansion of $\varepsilon' t$, the $L$-dependent corrections play a crucial role near the endpoints. Around $t\approx T/2$, the $L$-dependent corrections control the local speed and the number of excitations.

\begin{figure}
    \includegraphics[width=\columnwidth]{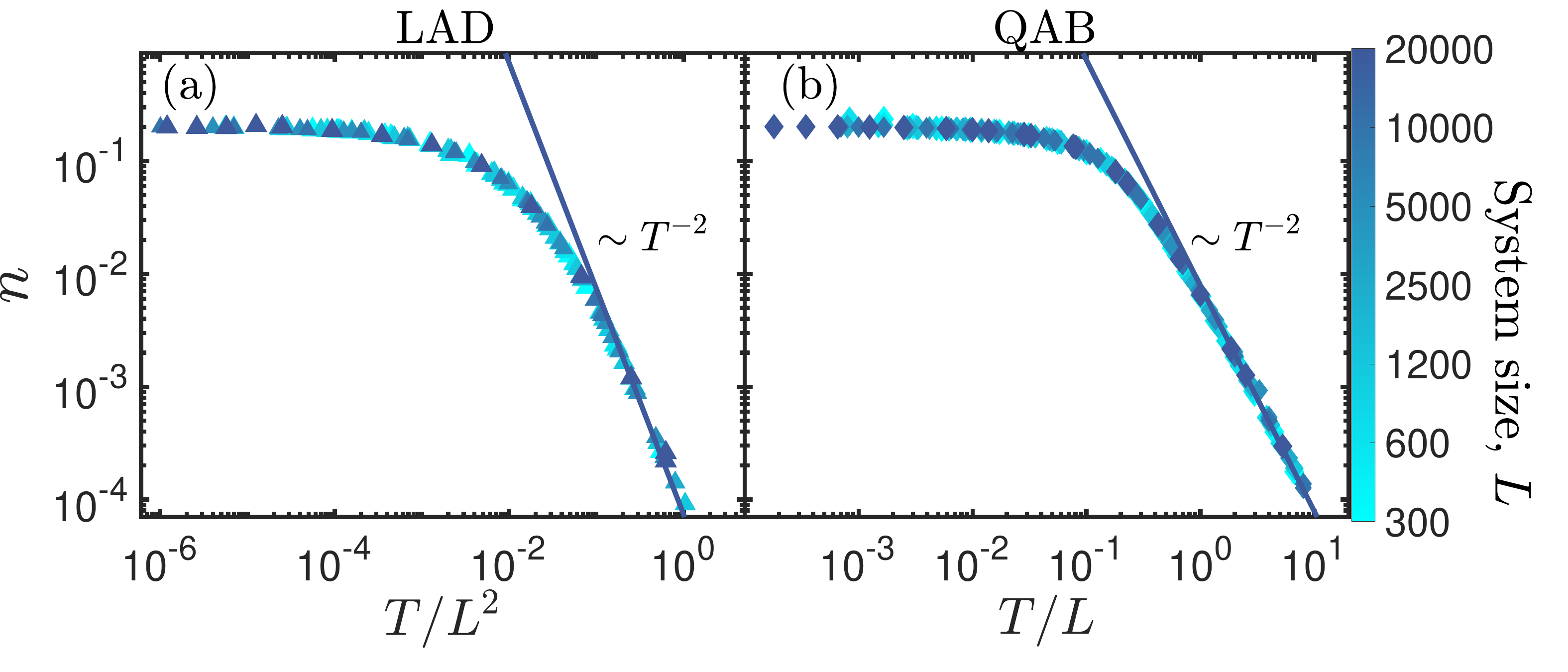}
    \caption{Scaling of the density of excitations $n$ for (a) LAD and (b) QAB. Both curves show a perfect collapse upon rescaling the evolution time $T$ with a power of the system size. The excitation densities exhibit a plateau with an extensive number of excitations, $n \sim \mathcal O(1)$. This regime grows extensively with $L$ and then turns immediately into an adiabatic-like decay $n \sim T^{-2}$. The universal KZ scaling regime is not present in either case.
    }
    \label{fig:TFIM_KZM_2panels}
\end{figure}

To characterize the efficiency of the LAD approach, the density of excitations $n$ is investigated as a function of $T$ in relation to the KZ mechanism. As shown in Fig.~\figpanel{fig:TFIM_KZM_2panels}{a}, $n$ exhibits a good scaling collapse for different values of $L$ as a function of $T/L^2$. In particular, for timescales $T_\mathrm{plateau} \lesssim L^2$ an extensive---in the thermodynamical sense, i.e.,\ linear in the system size---number of modes get excited with finite probability, as reflected by the $n\sim\mathcal O(1)$ plateau. After this scale, the density falls off as $\sim T^{-2}$. It follows that the time of onset of adiabaticity, defined by $n \approx 1/L$, scales as $T_\mathrm{ons} \sim T^{5/2}$. This is an important aspect that may be counter-intuitive: the time of onset of adiabaticity is longer than the adiabatic time itself, since $T_\mathrm{ad} \sim L^2$ for the linear schedules in the TFIM. The solution to the apparent paradox comes from realizing that $n$ takes contributions from all momentum modes, while LAD is built using the zero momentum mode and guarantees that only this mode is driven transitionlessly. Finally, it is interesting to note that the curve $n(T)$ obtained with the LAD procedure, Fig.~\figpanel{fig:TFIM_KZM_2panels}{a}, also lacks a universal KZ power-law scaling regime. 

The unusual features described above can be understood with a simple analytical argument based on the LZ formula. The curve $g_\mathrm{LAD}(t)$, Eq.~\eqref{eq:g_LAD}, is very flat except in a small region near the boundaries, where it drops very fast; see also Fig.~\ref{fig:LADprotocol}. Setting $\delta t = T/2-t$, one can expand Eq.~\eqref{eq:g_LAD} to lowest order in $1/L$ and $\delta t/T$, finding
\begin{equation}
    g_\mathrm{LAD}(t) \approx 1-\frac{1}{L}\frac{1}{\sqrt{\delta t/T}}.
\end{equation} 
The region of the fast drop is identified by $1-g_\mathrm{LAD}=\mathcal O(1)$, which sets $\delta t \sim T/L^2$. The breakdown of the plateau $n\sim \mathcal O(1)$ with an extensive number of excitations is identified via the condition $\dot g\sim\mathcal O(1)$. This implies the suppression of the $p_k$  for LZ transitions in all momentum modes. This is equivalent to requiring the finiteness of the fast drop region up to the leading order, $\delta t\sim\mathcal O(1)$.
Combining these conditions, one finds $T\sim L^2$, which sets the timescale at which an extensive number of momentum modes starts to be driven transitionlessly. Beyond this timescale, low-momentum avoided crossings are traversed with a typical local speed of $\dot g \sim 1/TL \sim L^{-3}$, while for the higher momenta, this velocity scales as $\sim 1/T \sim L^{-2}$. This latter case is similar in many aspects to the adiabatic regime of the linear schedules, explaining the observed decay of the defects $n \sim T^{-2}$.

\subsection{Quantum adiabatic brachistochrone (QAB)}
\label{sec:QAB}

In the previous section, it was argued that the LAD procedure fails to produce sensible annealing schedules $g(t)$. Slowing down at the avoided crossing of the zero mode can only be achieved at the expense of speeding up near the end of the evolution, exciting an extensive number of momentum modes. The QAB approach tries to solve this problem by weighting the adiabatic condition along all the paths with the norm of the derivative of the Hamiltonian, i.e., \ using the action $S_\mathrm{QAB}$, Eq.~\eqref{eq:QAB}, rather than applying it locally. When specified to the TFIM, the action reads
\begin{equation}
    \label{eq:S_QAB_TFIM}
    S_\mathrm{QAB} = L \int \mathrm dt \, \frac{\dot g^2(t)}{[1+g^2(t)-2g(t)\cos k_0]^2}\,,
\end{equation}
where again $k_0=\pi/L$ is the lowest momentum for a system of size $L$. Notice that the norm $\| \partial_t \hat{H} \|^2 = L\,\dot{g}^2$, as one can check in the fermionic basis (see Sec.~\ref{sec:TFIM_diagonalization}).
Despite equal weighting via the norm, the final expression is still governed by the zero mode energy gap. Therefore, the behavior is expected to be similar to that in the LAD approach discussed in Sec.~\ref{sec:LAD}.

The Euler-Lagrange equation for the action reads
\begin{equation}
    \label{eq:EL_QAB_TFIM}
    \der{}{t} \fd{S}{\dot g} = \fd{S}{g} \quad \implies\quad
    \ddot g = \frac{2\dot g^2(g-\cos k_0)}{1+g^2-2g\cos k_0}.
\end{equation}
As shown in App.~\ref{app:sec:QAB}, this equation can be integrated to
\begin{align}
    \label{eq:g_QAB}
    g_\mathrm{QAB}(t) = \cos k_0 - \sin k_0 \tan [\varepsilon(t-t_0)],
\end{align}
where $\varepsilon$ and $t_0$ are two integration constants with roles analogous to the ones in the LAD procedure. They have to be fixed with the boundary conditions $g(-T/2) = g_0 = 2$ and $g(T/2)=0$. At large system sizes, their expressions simplify to
\begin{equation}
    \varepsilon = \frac{1}{T} \left[\pi + \mathcal{O}\left(\frac{1}{L}\right) \right], \quad
    t_0 = T \left[\frac{\pi^2}{2L^3} + \mathcal{O}\left(\frac{1}{L^4}\right) \right].
\end{equation}

In order to investigate the efficiency of the QAB schedules, the scaling of the number of defects was computed numerically, again taking inspiration from the KZ studies. In agreement with the structure of the action in Eq.~\eqref{eq:S_eQAB_TFIM}, Fig.~\figpanel{fig:TFIM_KZM_2panels}{b} demonstrates that the defect density remains extensive up to a time $T_\text{plateau} \sim L$; then, a rapid decay takes place, with $n \sim T^{-2}$. It follows that the onset of adiabaticity, defined by $n(T_\mathrm{ons}) \approx 1/L$, takes a time $T_\mathrm{ons} \sim L^{3/2}$. Thus, the QAB approach does not provide efficient schedules $g_\mathrm{QAB}(t)$, as it is still governed by the local energy gap. Even though the zero mode can be driven adiabatically in a timescale $T \sim L$, this comes at the price of creating an extensive number of excitations at the avoided level crossing of higher momentum modes. Similarly to the LAD approach, the universal KZ regime does not emerge during the whole process. 

The explanation of the features noted above parallels that of the LAD procedure. The QAB schedule evolves with a vanishingly small slope, $\dot g\sim 1/TL$, for the dominating part of the process apart from the small regions around the endpoints $t=\pm T/2$. The size of this region is characterized by $\delta t=T/2-t$. Expanding the schedule in terms of $\delta t$ and $1/L$ yields
\begin{equation}
    1-g_\mathrm{QAB}(t) \approx \frac{1}{L}\tan\left(\frac{\pi}{2}-\frac{\delta t}{T}\right)
    \approx \frac{T}{\delta t L}.
\end{equation}
The fast drop regime is captured by the condition $1-g = \mathcal{O}(1)$, implying the scaling $\delta t \sim T/L$. Setting as well $\delta t = \mathcal{O}(1)$ sets the condition for the non-diverging local velocity $\dot{g}\sim\mathcal O(1)$ at the avoided crossings. This leads to $T \sim L$, identifying the timescale at which an extensive number of momentum modes are no longer excited. For longer times $T$, the local speed scales as $|\dot{g}|\sim 1/TL \sim L^{-2}$, and all transitions near the zero mode are suppressed. Avoided crossings at higher momenta $k\gtrsim L^{-1/2}$ are also traversed adiabatically with typical velocity $\sim 1/T \sim L^{-1}$. This leads to a regime similar to the case of the linear schedule beyond the onset of adiabaticity, thus verifying the scaling $n\sim T^{-2}$.

\subsection{Equivalence of  LAD and  QAB}
\label{sec:LAD=QAB}

Consider a time-dependent quantum Hamiltonian of the form
\begin{equation}
    \hat{H}(t) = \hat{H}_0 + g(t) \hat{H}_1,
\end{equation}
where $\hat{H}_0$ and $\hat{H}_1$ are time-independent Hermitian operators, and the only dependence on time is mediated by the control parameter $g(t)$. Hamiltonians of this form are the ones usually considered in most quantum control problems~\cite{Boscain2021Introduction,DAlessandro2021Introduction};  the TFIM in Eq.~\eqref{eq:H_TFIM}  falls in this category as well. Here, it is proven that for such Hamiltonians, LAD and the QAB may lead to the same control schedules $g(t)$. Indeed, it is common in the LAD literature to upper bound $\dot{g}(t)$ as 
\begin{equation}
    \label{eq:LAD_gdot_modified}
    \dot g(t) < \varepsilon \frac{[E_1(t)-E_0(t)]^2}{\| \ev*{\partial_g \hat{H}} \|},
\end{equation}
cf.\ Eq.~\eqref{eq:LAD_gdot}; this approximation was actually performed in the original reference~\cite{Roland2002Quantum}. On the other hand, the QAB can be modified as follows 
\begin{equation}
    \label{eq:QAB_modified}
    \tilde{S}_\mathrm{QAB} = \int \mathrm{d}t \, \frac{|\mel{E_0(t)}{\partial_t \hat{H}}{E_1(t)}|^2}{[E_1(t) - E_0(t)]^4},
\end{equation}
where the matrix element of the ground to the first excited state replaces the norm of $\partial_t \hat{H}$ in Eq.~\eqref{eq:QAB}. Here, it is proven that Eq.~\eqref{eq:LAD_gdot_modified} is the Euler-Lagrange equation for the QAB, Eq.~\eqref{eq:QAB}, while the Euler-Lagrange equation of the modified QAB action, Eq.~\eqref{eq:QAB_modified}, is exactly the LAD equation~\eqref{eq:LAD}.

The proof is straightforward. The Euler-Lagrange equation for the modified QAB action, Eq.~\eqref{eq:QAB_modified}, reads
\begin{equation}
    \der{}{t} \frac{2 \dot{g} |\mel{E_0}{\hat{H}_1}{E_1}|^2}{[E_1 - E_0]^4} = \dot{g}^2 \pd{}{g} \frac{|\mel{E_0}{\hat{H}_1}{E_1}|^2}{[E_1 - E_0]^4},
\end{equation}
from which
\begin{equation}
    2\ddot{g} \frac{|\mel{E_0}{\hat{H}_1}{E_1}|^2}{[E_1 - E_0]^4} = -\dot{g}^2 \pd{}{g} \frac{|\mel{E_0}{\hat{H}_1}{E_1}|^2}{[E_1 - E_0]^4}.
\end{equation}
Since the equation is autonomous, one can pose $\dot{g} \equiv p(g)$ and integrate it (see also App.~\ref{app:sec:QAB}), finding
\begin{equation}
    \dot{g} = p_0 \frac{[E_1 - E_0]^2}{|\mel{E_0}{\hat{H}_1}{E_1}|}\,,
\end{equation}
where $p_0$ is an integration constant. This equation is the same as the one defining LAD,  Eq.~\eqref{eq:LAD_gdot}. 

The proof of the other case proceeds in the exact same way; one only needs to replace $|\mel{E_0}{\hat{H}_1}{E_1}|$ with $\|\hat{H}_1\|$. 

The fact that LAD and the QAB lead to the same optimal schedules $g(t)$ is an interesting fact that---to the best of our knowledge---was noted only for the Grover search~\cite{Rezakhani2009Quantum}. It also sheds light on the role of $\varepsilon$ in LAD: an additional parameter is needed because the LAD condition, Eq.~\eqref{eq:LAD}, contains only first-order time derivatives, while one needs two free parameters to impose the endpoint boundary conditions at $t=-T/2$ and $t=T/2$. 

Notice that above it is essential that $g(t)$ be real; in the case of a complex control field, the equivalence of LAD and QAB seems not guaranteed. The same is true in the case of more control terms, e.g.,\ if $\hat{H}(t) = \hat{H}_0 + g_1(t) \hat{H}_1 + g_2(t)\hat{H}_2 + \cdots$: the QAB approach is more general since LAD in presence of more time-dependent driving functions seems ill-defined.

\section{Optimal control beyond the adiabatic condition}
\label{sec:opt_contr_beyond_adiab}

In Sec.~\ref{sec:opt_contr_adiab}, we considered optimal control strategies that take as starting point the adiabatic condition, i.e.,\ that require the drive to be slow with respect to the inverse gap with (Eq.~\eqref{eq:LAD}) or without (Eq.~\eqref{eq:EL_QAB_TFIM}) the transition matrix elements. In this Section, we introduce optimal control methods that go beyond the adiabatic condition, taking into account also the higher excited states. Specifically, in Sec.~\ref{sec:eQAB}, we extend the QAB to tame also transitions in excited states, while in Sec.~\ref{sec:NAQO}, we introduce a new action to be minimized, based on the avoided level crossing structure in the LZ formula. Finally, we compare our results with other well-known optimization schemes: the ONLP through a quantum critical point in Sec.~\ref{sec:ONLP} and the fully numerical CRAB optimization in Sec.~\ref{sec:CRAB}.

\subsection{Extended quantum adiabatic brachistrochrone}
\label{sec:eQAB}

A natural extension of the adiabatic optimization approaches is provided by involving in the QAB action the effect of all momentum modes:
\begin{equation}
    \label{eq:S_eQAB_TFIM}
    S_\mathrm{eQAB} = \sum_{k>0} \int \mathrm{d}t \, \frac{|\mel{\epsilon_{+,k}
    (t)}{\partial_t \hat{H}}{\epsilon_{-,k}(t)}|^2}{[2\epsilon_k(t)]^4},
\end{equation}
where $\ket{\epsilon_{\pm,k}(t)}$ denotes the ground ($-$) and excited ($+$) single-particle states in the $k$-th momentum mode, and $2\epsilon_k(t)$ equals the gap in the same mode. We call this action ``extended quantum adiabatic brachistochrone'' (eQAB). Notice that in the numerator of Eq.~\eqref{eq:S_eQAB_TFIM} there is the matrix element between the one-particle ground and excited states, rather than the norm $\| \partial_t \hat{H}\|$, as described in Sec.~\ref{sec:LAD=QAB}. This choice allows for more accurate handling of the transition rates in the independent TLSs.

It would be tempting to replace the summation over $k$ in Eq.~\eqref{eq:S_eQAB_TFIM} with an integral. However, this would lead to singularities in the equation for $g(t)$: if
\begin{align}
    S_\mathrm{eQAB} &\approx L \int \mathrm{d}t \int_0^\pi \frac{\mathrm{d}k}{2\pi}\, \frac{\dot{g}^2(t) \sin^2 k}{(1+g^2(t)-2g(t) \cos k)^3}  \\
    &\sim \int \mathrm{d}t \, \frac{\dot g^2(t)}{[1-g^2(t)]^3},
\end{align}
then the Euler-Lagrange equation reads
\begin{equation}
    \label{eq:eQAB_EL_nonregularized}
    \ddot g=\frac{3g\dot g^2}{1-g^2}.
\end{equation}
This equation is highly singular at $g=1$, i.e.,\ at the QPT. In particular, $g(t)$ is forced to slow down infinitely due to the continuum approximation before reaching $g=1$, and thus the QPT cannot be crossed in finite time. This drawback in the equation for $g(t)$ comes from letting $k$ reach 0. It is therefore important to retain the lower limit of integration to be $k_0=\pi/L$, which keeps track of the system size. 

In order to simplify the analytical expressions, a different
route is followed: $k$ is integrated between $0$ and $\pi$, but a regularization in the denominator of Eq.~\eqref{eq:S_eQAB_TFIM}, that shifts the singularity to $-k_0$, is introduced. This is achieved by 
\begin{equation}
    S_\mathrm{eQAB} \approx \int \mathrm{d}t \int_{-1}^1 \mathrm{d}x\, \frac{\sqrt{1-x^2}}{(1+ g^2 - 2gx + 2gk_0)^3},
\end{equation}
having changed variables to $x = \cos k$, expanded to lowest order in $1/L$, and left implicit the dependence on time for brevity. This approach captures all the essential analytical properties, as well as additional corrections in the numerator, while making only subleading errors. Performing the integral over $x$, one finds
\begin{equation}
    \label{eq:eQAB_regularized}
    S_\mathrm{eQAB} \approx \int\mathrm dt\,\frac{\dot g^2}{[(g^2-1)^2+4g(1+g^2)k_0]^{3/2}}.
\end{equation}

\begin{figure}
    \includegraphics[width=\columnwidth]{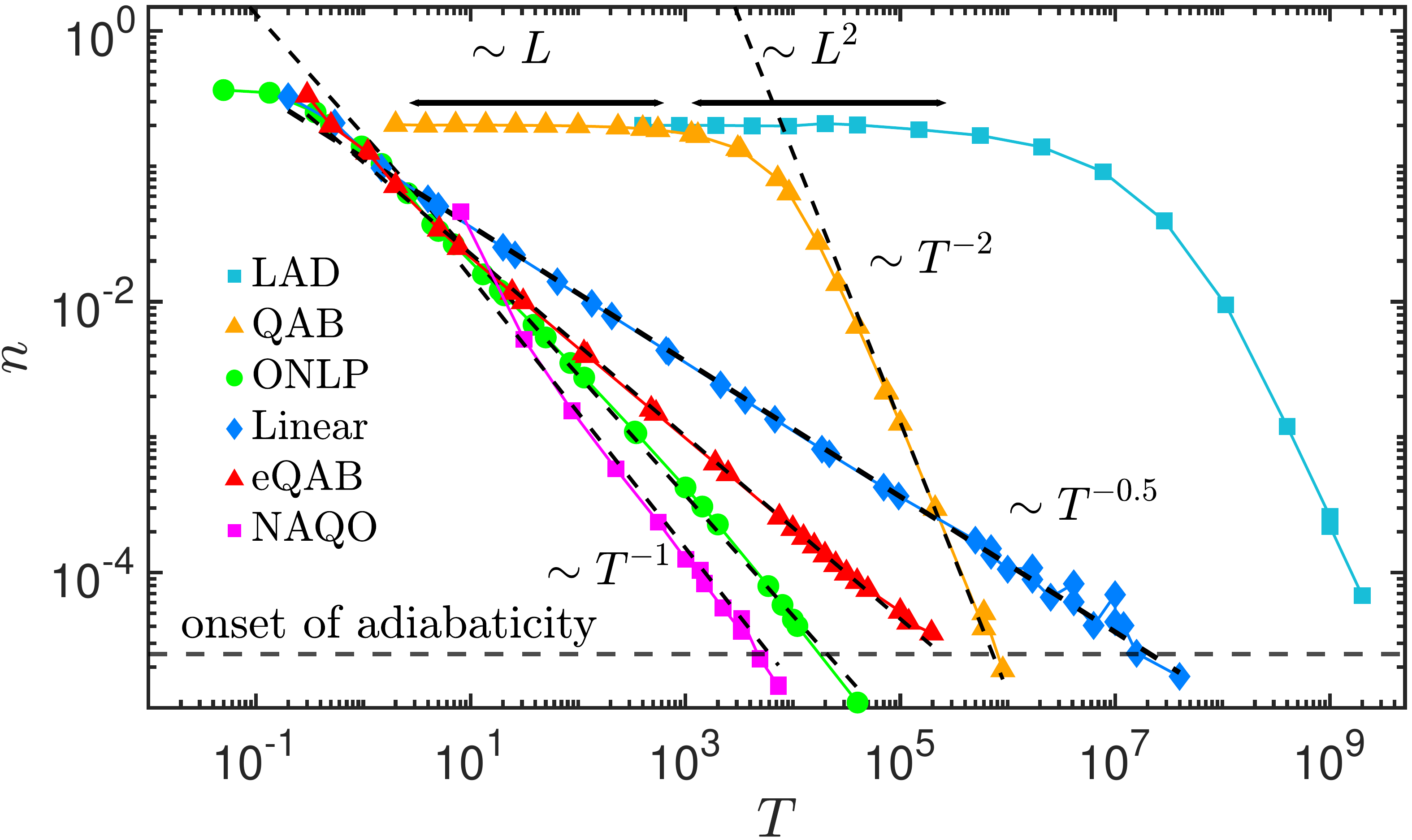}
    \caption{Scaling of the density of excitations $n$ as a function of the evolution time $T$ for the various schedules described in the main text. The system size is fixed to $L=40000$. The horizontal dashed line indicates the onset of adiabaticity, defined by $n\approx1/L$. The power laws obtained from the fits in the figure are summarized in Table~\ref{tab:summary}.}
    \label{fig:KZM_protocols}
\end{figure}

The minimization of the regularized eQAB action in Eq.~\eqref{eq:eQAB_regularized} with respect to $g$ leads to the differential equation
\begin{equation}
    \label{eq:eQAB_diff_eq}
    \ddot g = \frac{5\dot g^2 g(g^2-1)}{(g^2-1)^2+8k_0}.
\end{equation}
Details of the calculations are provided in App.~\ref{app:sec:eQAB}. Confronted with the non-regularized Eq.~\eqref{eq:eQAB_EL_nonregularized}, one can see that the $k_0$ term in the denominator keeps the differential equation regular and accounts for the gap dependence near the critical point.

In Fig.~\ref{fig:KZM_protocols}, the curve $n(T)$ obtained numerically is displayed. Remarkably, the timescale of the onset of adiabaticity does not improve with respect to the standard QAB, apart from an $\mathcal{O}(1)$ prefactor: $T_{\mathrm{ons}}\sim L^{3/2}$, as displayed in Fig.~\ref{fig:onset_scales}. It is interesting to notice that since the obtained $g$ schedules are independent of the system size, the universal KZ power-law regime is present without an $L$-dependent rapid quench plateau~\cite{Xia2024}. The corresponding decay of the defects follows the approximate power law $n\sim T^{-0.66}$, as found from the fit in Fig.~\ref{fig:KZM_protocols}. 

The reason why the eQAB approach does not provide a significant improvement over the QAB  can be understood as follows. Equation~\eqref{eq:S_eQAB_TFIM} is an unweighted average over all momentum modes of an action of the QAB form. There are two consequences to this fact. First, the zero-momentum mode still dominates, as exemplified by the Euler-Lagrange equation becoming singular in the thermodynamic limit. Second, the main contribution to the excitations arises from the vanishingly small regions around the avoided crossings. However, the action in Eq.~\eqref{eq:S_eQAB_TFIM} does not weight properly these short time intervals where the gaps are closing, masking their true impact. In addition, the non-adiabatic rates within the eQAB action do not account accurately for excitation probabilities around the avoided crossings. In other words, the LZ transitions are insensitive to the local adiabatic conditions when the latter is shorter than the global adiabatic time-scale, $T<T_\mathrm{ad}$. Increasing the time intervals where the local adiabatic conditions are taken into account misses the source of non-adiabaticity.
These two facts lead to a considerably weaker effective condition to suppress excitations.

\subsection{Non-Adiabatic Quantum Optimization (NAQO) of Landau-Zener transitions}
\label{sec:NAQO}

Above, it was argued that tailoring control schedules by the local adiabatic condition, Eq.~\eqref{eq:T_ad}, leads to suboptimal results, as higher-excited modes are simply not considered. Remarkably, no relevant improvement in the timescale of the onset of adiabaticity was found even by the simultaneous optimization of all the momentum modes with the eQAB, see Sec.~\ref{sec:eQAB}. These results indicate that even in simple cases, many-body effects lead to non-trivial behavior, deteriorating the efficiency of the conventional optimal control strategies.

In this section, we show that it is possible to design better schedules by focusing attention on the main source of non-adiabatic effects, i.e., the small neighborhood of the avoided level crossings. As anticipated at the end of the last section, the schedule $g(t)$ needs to go slow only when there is a high probability of exciting a momentum mode, i.e.,\ near an avoided crossing. This is exactly where the LZ formula applies, both in the KZ scaling regime and beyond the onset of adiabaticity~\cite{Dziarmaga2005Dynamics,DamskiZurek06,Defenu_LRK,Defenu_LMG,Zurek2005Dynamics}. Designing efficient control schedules requires the simultaneous suppression of LZ transitions in both low and high-momentum states. To capture this in an analytical way, the following action is introduced:
\begin{equation}
    \label{eq:LZ_action_concept}
    S_\mathrm{NAQO} = \sum_k \exp\left[-\frac{2\pi k^2}{\dot g_c(k)}\right],
\end{equation}
where $g_c(k) := g(t_c(k))$ is the position of the avoided crossing as a function of $k$; see Eq.~\eqref{eq:AvCross_pos}. We denote Eq.~\eqref{eq:LZ_action_concept} as the NAQO action. 
This action can also be understood as refining the uniform averaging over time in Eq.~\eqref{eq:EL_QAB_TFIM} to only those time instances where the $k$-th avoided crossing is formed, fixed by the relation $g(t^*)=\cos k$, Eq.~\eqref{eq:AvCross_pos}. Additionally, the integrand is modified by using the LZ formula instead of the local-adiabatic transition rate.
Notice that $S_\mathrm{NAQO}$ is defined only for values of $g<1$, i.e.,\ where there actually are avoided crossings. How to extend the action for $g>1$ is explained below.

In order to get to a self-consistent expression, one can approximate
\begin{equation}
    S_\mathrm{NAQO} \approx \frac{L}{2\pi} \int_0^{\pi/2} \mathrm dk\, \exp[- \frac{2\pi (1-g^2_c(k))}{\dot{g}_c(k)}].
\end{equation}
The integral runs only up to $\pi/2$ since higher momenta do not display avoided crossings. Restricting to monotonic schedules with $\dot{g} \leq 0$, one can change the integration variable from $k$ to the time $t$, arriving at
\begin{equation}
    \label{eq:LZ_action}
    S_\mathrm{NAQO} \sim \int \mathrm dt\; \frac{\dot{g} \, e^{-2\pi (1-g^2)/\dot g}}{\sqrt{1-g^2}},
\end{equation}
where the dependence of $g$ on time is omitted for brevity. The Euler-Lagrange equation reads
\begin{equation}
    \label{eq:LZ_opt_diffeq}
    4\pi(1-g^2)g\dot g^2+2\pi(1-g^2)^2\ddot g-g\dot g^3=0.
\end{equation}
As in the eQAB, this equation is singular near $g=1$.
However, instead of regularizing it by shifting the singularity in the action, it is convenient to adjust the initial condition to the position of the LZ transition in the zero mode. In particular, the schedule is considered to start from $g_c(0) = \cos (k_0) \approx 1-k^2_0/2$. 
Note that the solution close to this point is given by
\begin{equation}
    g(t)\approx W \left(-\pi^2 e^{8\pi t}/2L^2 \right)+1 \qquad \text{for } t\ll T,
\end{equation}
where $W(t)$ denotes the Lambert function, defined as the principal solution of 
\begin{equation}
    \label{eq:Lambert}
    W e^W=t. 
\end{equation}
This leading order analysis also reveals that the initial derivative starts as $\dot g(0) \sim L^{-2}$. The second boundary condition, namely $g(T/2)=0$, fixes the shape of $g(t)$ for $t \in [0,T/2]$; the function $g(t)$ for $t \in [-T/2,0]$ is obtained by symmetrization~\footnote{Actually, the numerical solution and the extraction of optimal schedules are obtained via a shooting method, parametrizing the solutions by the initial slope, $\dot g(0)=sL^{-2}$ with $s$ a free parameter. As an initial condition, this slope determines numerically the final time $T$, allowing for the analysis in terms of $T$ and $L$.}. Details of the calculation are presented in App.~\ref{app:sec:NAQO}.

As demonstrated in Figs.~\ref{fig:KZM_protocols} and \ref{fig:onset_scales}, the schedules $g(t)$ obtained via the NAQO action represent a significant improvement over the LAD, QAB and eQAB ones. The adiabatic timescale is reduced to that required 
for transitionless passage through the zero mode in schedules based on the adiabatic condition. There curves $n(T)$ display an $L$-independent fast-quench plateau, after which $n\sim T^{-1}$. The timescale of onset of adiabaticity is found to be $T_\mathrm{ons} \sim L$.

\begin{table}
    \centering
    \begin{tabular}{|c|c|c|c|} \hline
        schedule       &plateau        &decay            &$T_\mathrm{ons}$ \\ \hhline{|=|=|=|=|}
        linear  &$\sim L^0$     &$\sim T^{-1/2}$  &$\sim L^{2}$ \\ \hline
        LAD     &$\sim L^2$     &$\sim (T/L^2)^{-2}$    &$\sim L^{5/2}$ \\ \hline
        QAB     &$\sim L$       &$\sim (T/L)^{-2}$    &$\sim L^{3/2}$ \\ \hline
        eQAB    &$\sim L^0$     &$\sim T^{-0.66}$ &$\sim L^{1.5}$ \\ \hline
        NAQO     &$\sim L^0$     &$\sim T^{-1}$    &$\sim L$ \\ \hline
        ONLP    &$\sim L^0$     &$\sim T^{-0.84}$    &$\sim L^{1.2}$ \\ \hline
        average CRAB    &$\sim L^0$     &$\sim T^{-0.7}$    &$\sim L^{1.43}$ \\ \hline
        best CRAB    &$\sim L^0$     &$\sim T^{-0.8}$    &$\sim L^{1.24}$ \\ \hline
    \end{tabular}
    \caption{The density of excitations $n(T)$ displays a plateau and then a drop, as shown in Fig.~\ref{fig:KZM_protocols}. Here, the scaling of these two regimes is summarized, together with the resulting timescale of onset of adiabaticity $T_\mathrm{ons}$. The LAD is described in Sec.~\ref{sec:LAD}, QAB in Sec.~\ref{sec:QAB}, the eQAB in Sec.~\ref{sec:eQAB}, the NAQO in Sec.~\ref{sec:NAQO}, the ONLP in Sec.~\ref{sec:ONLP}, the CRAB algorithm in Sec.~\ref{sec:CRAB}.}
    \label{tab:summary}
\end{table}

\subsection{Optimal non-linear passage}
\label{sec:ONLP}

\begin{figure}
    \includegraphics[width=.48\textwidth]{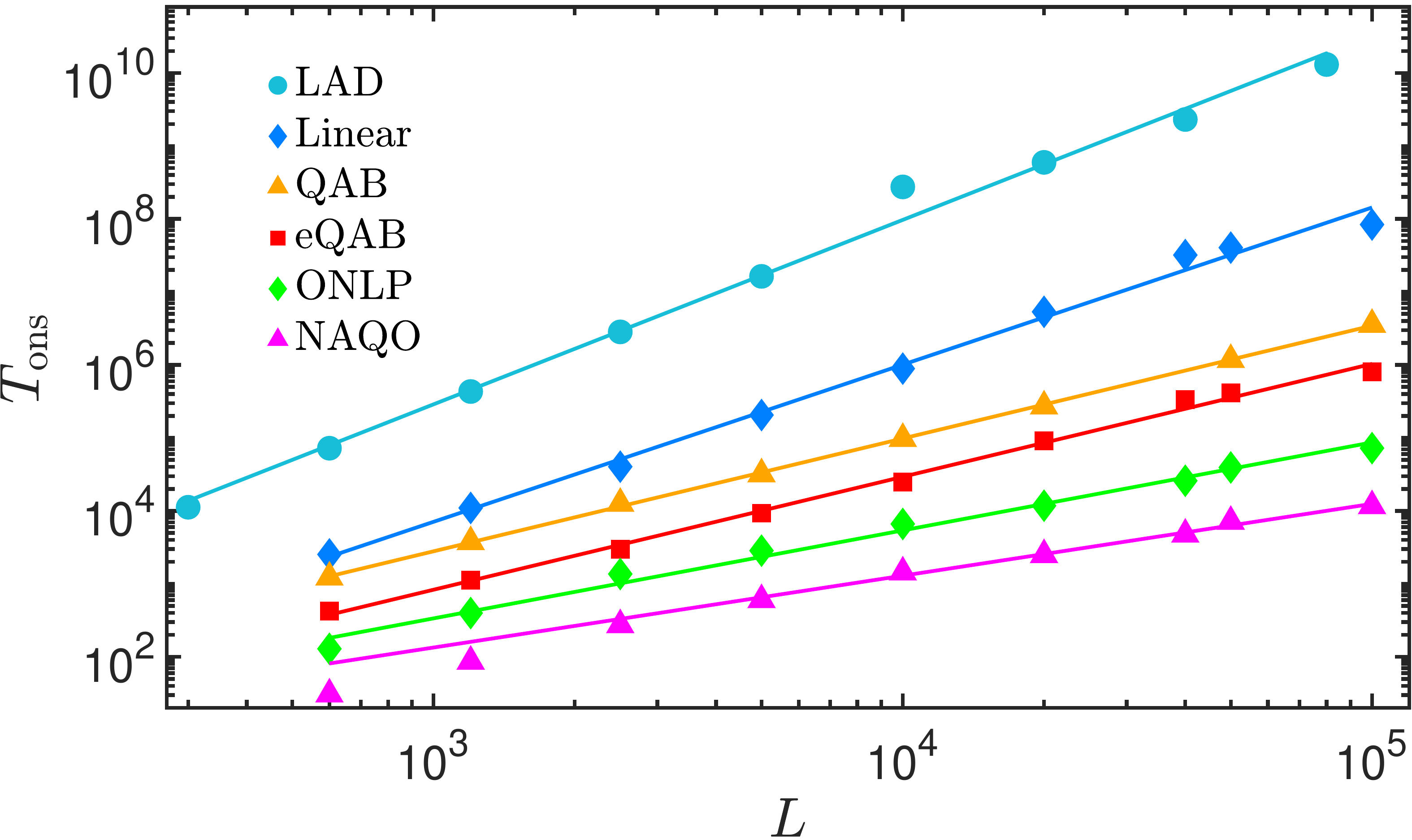}
    \caption{Timescale required to reach the onset of adiabaticity, defined by $n(T_\mathrm{ons}) = \mathcal{O}(1)$. The values of $T_\mathrm{ons}$ are extracted from Fig.~\ref{fig:KZM_protocols}. The fits $T_\mathrm{ons} = T_\mathrm{ons}(L)$ provide power laws compatible with the ones from the fits $n = n(T)$, as can be seen from Table~\ref{tab:summary}.}
    \label{fig:onset_scales}
\end{figure}

For the sake of completeness, we compare our results to the performance of optimized power-law schedules that were obtained in Refs.~\cite{Barankov08,Diptiman08}:
\begin{equation}
    g(t)=1-\left\lvert\frac{2t}{T}\right\rvert^r\mathrm{sign}(t).
\end{equation}
Within this approach, the KZ exponent $\alpha_\mathrm{KZ}$ is maximized as a function of $r$ in order to determine the optimal schedule once the universality class of the transition is known. The solution is 
\begin{equation}
    r=\ln\left[\frac{1}{TC}\ln(TC)\right],\qquad C\approx 14.6.
\end{equation}
As demonstrated in Figs.~\ref{fig:KZM_protocols} and \ref{fig:onset_scales}, this schedule provides a faster KZ decay and shorter $T_\mathrm{ons}$ if compared to LAD and the QAB, but a slower KZ decay and longer $T_\mathrm{ons}$ compared to the NAQO. The improvement of the adiabatic time-scale compared to $T_\mathrm{ons}\sim L^{1.5}$ is the consequence of the fact the ONLP procedure does account for higher momentum states. This is encoded in the $z$ exponent via the relation of $\epsilon_k\sim k^z$. However, owing to the leading order structure, it still fails to accurately capture all the higher momentum modes, which explains its reduced performance with respect to the NAQO.

In Ref.~\cite{Wu2015Optimal}, additional numerical optimization for the exponent $r$ was employed; however,  no improvement was observed for the accessible system sizes.

\subsection{Fully numerical optimization}
\label{sec:CRAB}

Above, several semi-analytical optimization strategies for crossing a second-order QPT were considered. Here, the results obtained are benchmarked against a purely numerical approach, i.e.,\ the CRAB algorithm for quantum optimal control~\cite{Doria11,Caneva11,Muller2022One}. 

The idea of the CRAB algorithm is very simple. The control field $g(t)$ is expanded in a basis of functions $\{\varphi_m(t)\}$ on the interval $[-T/2,T/2]$ as 
\begin{equation}
    g(t) = \sum_{m=0}^\infty g_m \varphi_m(t).
\end{equation}
Then, only a handful of $\varphi_m$ are retained in the expansion (e.g.,\ fast oscillating functions are discarded), and the coefficients $g_m$ are optimized iteratively with a gradient-free optimization algorithm, using as a cost function the fidelity to the target state. 

Here, a few modifications are made for convenience. First, the series expansion is fixed to
\begin{equation}
    g(t) = g_0 \left( \frac{1}{2} - \frac{t}{T} \right) \left[ 1+ \sum_{m=1}^{m_{\max}} g_m T_{2m}\left(\frac{2t}{T}\right) - \sum_{m=1}^{m_{\max}} g_m  \right],
\end{equation}
where $T_m$ are the Chebyshev polynomials of the first kind. The form chosen above ensures that the expansion is made around the linear schedule (when all $g_m = 0$) and that $g(-T/2)=g_0$ and $g(T/2)=0$. Second, instead of the fidelity, the final excitation density $n$ is used as a cost function. The chosen optimization method was an adaptive Nelder-Mead's simplex method~\cite{Gao2012Implementing,SciPy}.

\begin{figure}
    \includegraphics[width=\columnwidth]{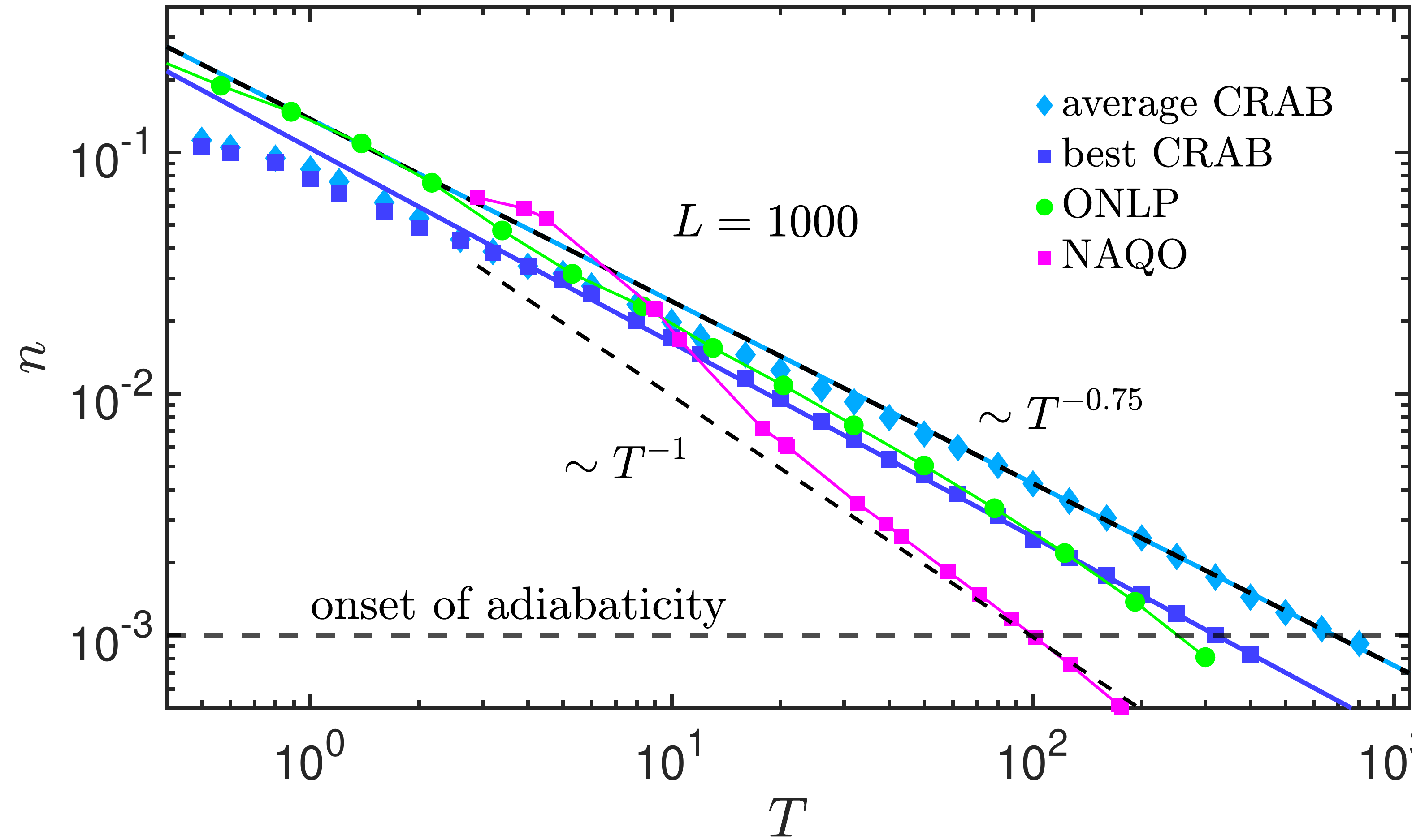}
    \caption{Comparison of the analytical NAQO and ONLP with the average and best CRAB numerical optimization. The system size is fixed with $L=1000$ since the CRAB algorithm becomes numerically demanding for larger systems. While, on average, the CRAB search does not find good schedules, the best instance over $1000$ trials falls near the ONLP results. The NAQO results show an anomalously fast decay here due to finite-size effects, while they approach $n\sim T^{-1}$ at larger sizes, see Fig.~\ref{fig:KZM_protocols}.}
    \label{fig:TFIM_CRAB}
\end{figure}

The results of the CRAB optimization are displayed in Fig.~\ref{fig:TFIM_CRAB}. In order to present a fair comparison between the CRAB and the ONLP and NAQO schemes, it was chosen not to give the numerical optimization overwhelming resources. Thus, it was fixed $m_{\max} = 5$, and the search for the optimum was run 1000 times for each instance of system size $L$ and final time $T$, starting from a random sequence of $g_m$ uniformly distributed over $[-0.1,0.1]^{m_{\max}}$. We checked that running the optimum search 250 and 500 times did not change appreciably the results. One can see that, under these conditions, the solutions found by CRAB perform slightly worse than the NAQO both on average and for the best minimum found overall. Furthermore, the accessible system sizes are much smaller, given that the optimization procedure is numerically demanding: we experienced that obtaining a NAQO schedule takes order of few seconds on a 2.4 GHz Intel Core i5 MacBookPro, and this time is independent of the system size $L$, while a single CRAB optimum search requires simulating the time evolution for a system of size $L$ for each cost function call, with a total time reaching $\sim 12$ hours on the same machine at $L=1000$.

\begin{figure}
    \includegraphics[width=.48\textwidth]{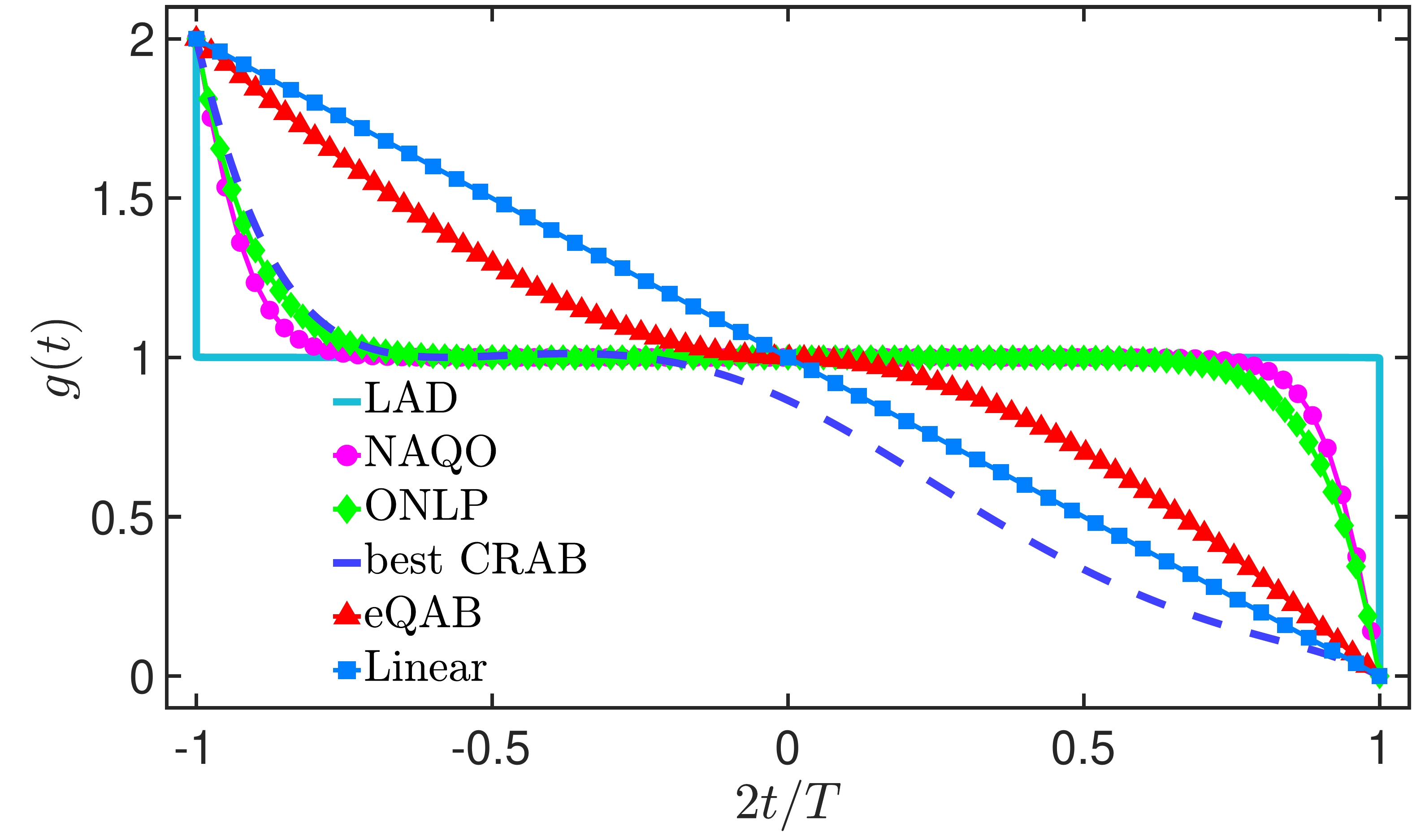}
    \caption{Optimal control schedules for $L=40000$ and $T=5000$. The LAD scheme provides schedules that become singular already at large but finite $L$, being composed of a flat, very slow crossing of the critical point and two sudden jumps at the endpoints. Such jumps excite a dominant fraction of momentum modes, yielding poor results. The extended quantum adiabatic brachistochrone (eQAB) is closer to the other extreme represented by the linear schedule, and thus, it does not properly suppress the excitations either. The ONLP and the NAQO are halfway in between and comparable to each other to the naked eye. The CRAB schedule follows a similar structure as the ONLP and the NAQO but exhibits further non-monotonic regions.}
    \label{fig:protocols}
\end{figure}

Finally, let us illustrate the shape of the considered optimal control protocols in Fig.~\ref{fig:protocols}.
Interestingly, the typical shape of the extended QAB schedules is of the opposite nature to those of the LAD and QAB approaches. These latter are extremely flat around $g=1$, with a slope scaling as $\sim (TL)^{-1}$, and present a sharp jump around $t=\pm T/2$. The former is more similar to the linear schedule, without dramatic changes in the local velocity $\dot g(t)$. The ONLP and NAQO protocols exhibit features are of an intermediate structure, as they tend to change slowly around the critical point, however do not speed up significantly at high momenta either. However, the NAQO schedule exhibits a more optimal shape balancing the simultaneous effects of both low and high momentum states. In contrast, the CRAB approach exhibits additional features and a surprising non-monotonic behavior close to the QPT.

\section{Extension to long-range Kitaev models}
\label{sec:LRK}

In this section, the analytical NAQO framework is extended to the long-range Kitaev model (LRKM).

\subsection{Model and diagonalization}
\label{sec:LRK_diagonalization}

The LRKM describes a $p$-wave superconductor of spinless free fermions. In the case of power-law-decaying hopping and $p$-wave-type pairing interactions, the Hamiltonian reads (with the energy scale $J$ fixed again to 1)~\cite{Vodola_LRK,Dutta_LRK,Defenu_LRK}
\begin{multline}
    H(t)= -g(t) \sum_{i=1}^L c^\dagger_i c_i \\
    -\sum_{i=1}^L \sum_{r=1}^{L/2} \left[ j_{r,\alpha} c^\dagger_ic_{i+r} + d_{r,\beta} c^\dagger_{i} c^\dagger_{i+r} +\,\mathrm{h.c.} \right],
\end{multline}
where $c^\dagger_i$ and $c_i$ are fermionic creation and annihilation operators. Above, the interaction ranges over half of the chain in order to avoid double counting. The strength of the couplings is given by
\begin{equation}
    j_{r,\alpha}=\frac{1}{N_\alpha} r^{-\alpha}, \qquad
    d_{r,\beta}=\frac{1}{N_\beta} r^{-\beta},
\end{equation}
where $\alpha,\beta>1$ to ensure that the energy spectrum is bounded. Consequently, the normalizations are finite~\cite{LR_systems_book}:
\begin{equation}
    N_\gamma = 2\sum_{r=1}^{L/2}r^{-\gamma}
    \stackrel{L\to \infty}{\approx} 2\zeta(\gamma), \quad \text{for } \gamma=\alpha,\beta.
\end{equation}
Above, $\zeta(s) = \sum_{n=1}^\infty n^{-s}$ is the Riemann zeta function. With these exponents, the LRK model exhibits a second-order phase transition at $g_c=2$~\cite{Kitaev_Majorona_LR_2001}. The corresponding order parameter is provided by the bulk topological invariant $w$, counting the number of the doubly degenerate Majorana zero modes (MZM). Inside the trivial phase, $g>g_c$, the ferromagnetic ground state is non-degenerate with $w=0$ MZMs. For $g<g_c$, two degenerate MZMs are exponentially localized at the ends of the chain in the case of open boundary conditions, $w=1$~\cite{Albrecht2016MZM_edgestates, MZMmodes_Science_20212}. The phase boundary is invariant under variation of the long-range exponents $\alpha,\,\beta>1$ ~\cite{LongRangePowerLawSC_Delgado_2017, LRK_powerlawDelAnna2017}. The energy spectrum at finite $L$ is gapped at $g=g_c$, but as $L\to\infty$ the gap closes as a power law (see Eq.~\eqref{eq:gap_LRKM} below), giving rise to the breakdown of adiabaticity and excitation formation upon crossing the critical point~\cite{Vodola_LRK,Dutta_LRK,Defenu_LRK,Balducci2023Large}.

Before proceeding to the dynamical properties, the diagonalization procedure of the LRK models is briefly summarized. Employing the Fourier representation of the fermionic creation and annihilation operators, $\hat c_k=e^{i\pi/4}\,\sum_{r=-\infty}^\infty\,\hat c_r\,e^{ikr}/\sqrt L$, the Hamiltonian decomposes into the direct sum of independent TLSs for each momentum $k=\frac{\pi}{L},\,\frac{2\pi}{L},\,\dots\,\pi$: 
\begin{align}
	H &= 2 \sum_{k>0} \hat{\psi}_k^\dagger \left[ (g(t)/2-j_\alpha(k))\tau^z + d_\beta(k) \, \tau^x \right] \hat{\psi}_k \\
    &\equiv 2 \sum_{k>0} \hat{\psi}_k^\dagger H_k(t) \hat{\psi}_k,
\end{align}
where the momentum-space spinor representation was employed: $\hat{\psi}_k := (\hat{c}_k, \hat{c}_{-k}^\dagger)^T$. Except for the different expression of $H_k$, this is the same structure encountered in the TFIM; see Eq.~\eqref{eq:H_TFIM_momentum}. Here, the momentum-dependent diagonal and off-diagonal terms are given by the Fourier transforms of the pairing and hopping~\cite{Dutta_LRK,Defenu_LRK}:
\begin{align}
    &j_\alpha(k)=N^{-1}_\alpha\sum_{r=1}^{L/2}\,r^{-\alpha}\,\cos(kr)\approx \mathrm{Re}\left[\mathrm{Li}_\alpha(e^{ik})\right],\\
    &d_\beta(k)=N^{-1}_\beta\sum_{r=1}^{L/2}\,r^{-\beta}\,\sin(kr)\approx \mathrm{Im}\left[\mathrm{Li}_\beta(e^{ik})\right].
\end{align}
The corresponding single-particle energy spectrum is given by
\begin{equation}
    \label{eq:LRK_SP_energies}
    \epsilon_{\pm,k}(t)=\pm 2 \sqrt{\big[g(t)/2-j_\alpha(k)\big]^2+d^2_\beta(k)}.
\end{equation}
As for the TFIM, these TLSs exhibit avoided level crossings at the local minima of the energies in Eq.~\eqref{eq:LRK_SP_energies}. In particular, it was shown in Ref.~\cite{Defenu_LRK} that the position of the gaps, their size dependence, and the associated critical exponents can be divided into four different sectors according to the $\alpha,\beta$ exponents.

The position $g_c(k)$ of the avoided crossing in the $k$-th mode is a complicated expression of the momentum and the $\alpha$ exponent. However, the main analytical characteristics can be extracted from their leading order behavior:
\begin{equation}
    g_c(k)\approx
    \begin{cases}
        2+2\sin(\alpha\pi/2)\frac{\Gamma(1-\alpha)}{\zeta(\alpha)}k^{\alpha-1} &\alpha<3\\
        2-\frac{\zeta(2-\alpha)}{\zeta(\alpha)}k^2 &\alpha>3.
    \end{cases}
\end{equation}
The complete expressions are summarized in App.~\ref{app:LRK}.
Similarly, the energy gap minima $d_\beta(k)$ at the avoided crossings are investigated by the leading order behavior in the momentum, depending only on the $\beta$ exponent:
\begin{equation}
    \label{eq:gap_LRKM}
    d_\beta(k)\approx
    \begin{cases}
        \cos(\beta\pi/2)\frac{\Gamma(1-\beta)}{\zeta(\beta)}k^{\beta-1} &\beta<2\\
        \frac{\zeta(\beta-1)}{\zeta(\beta)}k    &\beta>2.
    \end{cases}
\end{equation}
Marginal cases with additional logarithmic momentum dependences at $\alpha=3$ and $\beta=2$~\cite{Defenu_LRK} are not addressed in this work. 

The ground state gap in the thermodynamic limit always scales with the control parameter as $\sim \lvert g-g_c\rvert^{-1}$ fixing $z\nu=1$, whereas both $z$ and $\nu$ strongly depend on $\alpha$ and $\beta$. These relations are obtained from the ground state energy gap at the critical point via the leading-order scaling $\sim k^z$. As the gap is two times the single-particle energies, Eq.~\eqref{eq:LRK_SP_energies}, only the leading order behavior of the latter is investigated  at $g=g_c=2$:
\begin{equation}
    \label{eq:Gap_relations}
    \epsilon_k(g_c) \sim
    \begin{cases}
        k^{\min\{\alpha,\beta\}-1} &\alpha<3,\,\beta<2\\
        k^{\beta-1}                &\alpha>3,\,\beta<2\\
        k^{\alpha-1}               &\alpha<3,\,\beta>2\\
        k                          &\alpha>3,\,\beta>2,
    \end{cases}
\end{equation}
which fixes the critical exponents to
\begin{equation}
    z =
    \begin{cases}
        \min\{\alpha,\beta\}-1  &\alpha<3,\,\beta<2\\
        \beta-1                 &\alpha>3,\,\beta<2\\
        \alpha-1                &\alpha<3,\,\beta>2\\
        1                       &\alpha>3,\,\beta>2,
    \end{cases}
\end{equation}
and $\nu=1/z$. The fourth sector ($\alpha>3$, $\beta>2$) resembles in many aspects the TFIM with identical equilibrium properties in the limit $\alpha,\beta\rightarrow\infty$. 

\begin{figure}[t]
    \centering
    \begin{tikzpicture}[scale=1.4]
        \draw (0,0) -- (5,0);
        \draw (0,0) -- (0,4);
        
        \draw (0,0) rectangle (5,4);
        
        \draw[dashed] (3,0) -- (3,4);
        \draw[dashed] (0,2) -- (5,2);

        \draw[thick] (0,0) -- (2,2);
        \draw[thick] (2,2) -- (2,4);
        
        \node at (2.5,-0.8) {\rotatebox{0}{$\alpha$}};
        \node at (-0.8,2) {\rotatebox{90}{$\beta$}};

        \node at (.9,2.25) {\rotatebox{60}{Dynamical power-law scaling}};

        \node at (3.15,1.9) {\rotatebox{60}{\large Kibble-Zurek scaling}};

        \node at (1.25,4.25) {\rotatebox{0}{Long range pairing}};

        \node at (4,4.25) {\rotatebox{0}{Short range pairing}};

        \node at (5.25,.95) {\rotatebox{270}{Long range hopping}};

        \node at (5.25,3.15) {\rotatebox{270}{Short range hopping}};

        \foreach \x in {0,1,2,3,4,5} {
            \draw (\x,0) -- (\x,-0.1) node[anchor=north] {\x};
        }
        \foreach \y in {0,1,2,3,4} {
            \draw (0,\y) -- (-0.1,\y) node[anchor=east] {\y};
        }
    \end{tikzpicture}
    \caption{Schematic diagram of the KZ and dynamical power-law-scaling regimes in the LRKM as a function of $\beta$ and $\alpha$. Excitation formation is governed by the KZ scaling for $\alpha>2$ and $\beta<\alpha<2$. The dynamical power-law-scaling phase emerges for $\alpha<2$}
    \label{fig:LRK_regimes}
\end{figure}
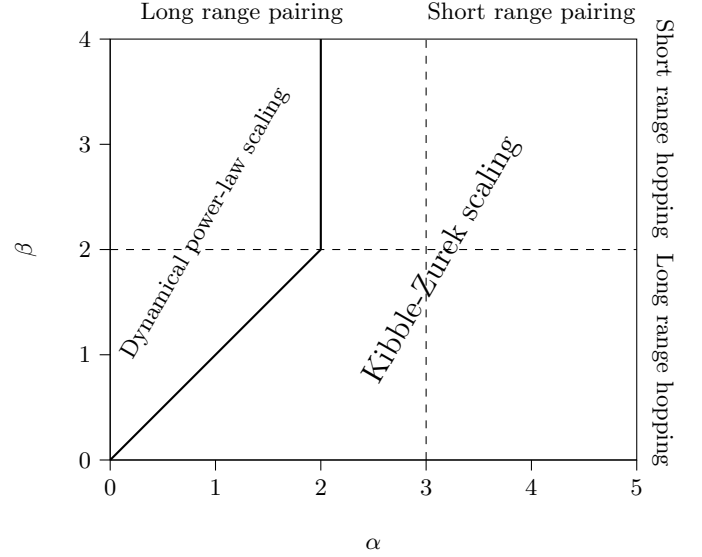

As a result of these properties, LRK models can be divided into four different sectors based on the values of $\alpha$ and $\beta$; see also Fig.~\ref{fig:LRK_regimes}. However, from the point of view of dynamics, the sectors to consider are slightly different and boil down to two power-law-scaling regimes, as discussed thoroughly in Ref.~\cite{Defenu_LRK}. In the short-range pairing sector, $\alpha>3$, excitations formation is governed by the universal KZ scaling law. The behavior changes drastically for strong long-range pairings, $\alpha<\mathrm{\min}\{\beta,2\}$, and a universal dynamical power-law-scaling regime arises that is not captured by the KZ mechanism. In the following subsections, the NAQO method will be tested separately in the two dynamical regimes.
 
\subsection{Optimal control in the universal KZ scaling regime}
\label{sec:LRK_KZregime}

When crossing the LRKM quantum critical point with a linear ramp given by
\begin{equation}\label{eq:lienarrampLRK}
    g(t)=2\left(1-\frac{2t}{T}\right), \quad t\in[-T/2,T/2],
\end{equation}
the number of topological defects above the final ground state decays according to universal power laws of $T$ \cite{Dutta_LRK,Defenu_LRK,Balducci2023Large,Singh23}. The sum of LZ transitions in the momentum modes captures the average defect generation. 

In the KZ scaling regime, identified by $1<\beta<\alpha<2$ and $\{\beta>1,\,\alpha>2\}$, the LZ excitation picture is in good agreement with the universal KZ power-law scaling described by the equilibrium critical exponents, $z$ and $\nu$. In particular, in the short-range hopping regime, $\beta>2$ ($\alpha>2$ is understood to be in the KZ scaling regime), the defect density scales as
\begin{equation}
    n\sim T^{\,-1/2}.
\end{equation}
In the long-range hopping sector, $\{\beta<\min(2,\alpha),\,\alpha<3\}$ and $\{\beta<2,\,\alpha>3\}$ for long- and short-range pairings, respectively, the decay becomes faster:
\begin{equation}
    n\sim T^{\,-1/(2\beta -2)}.
\end{equation}
In both cases, the exponents are equal to the KZ predictions of $T^{-\nu/(1+z\nu)}$.

\begin{figure}    
    \includegraphics[width=0.48\textwidth]{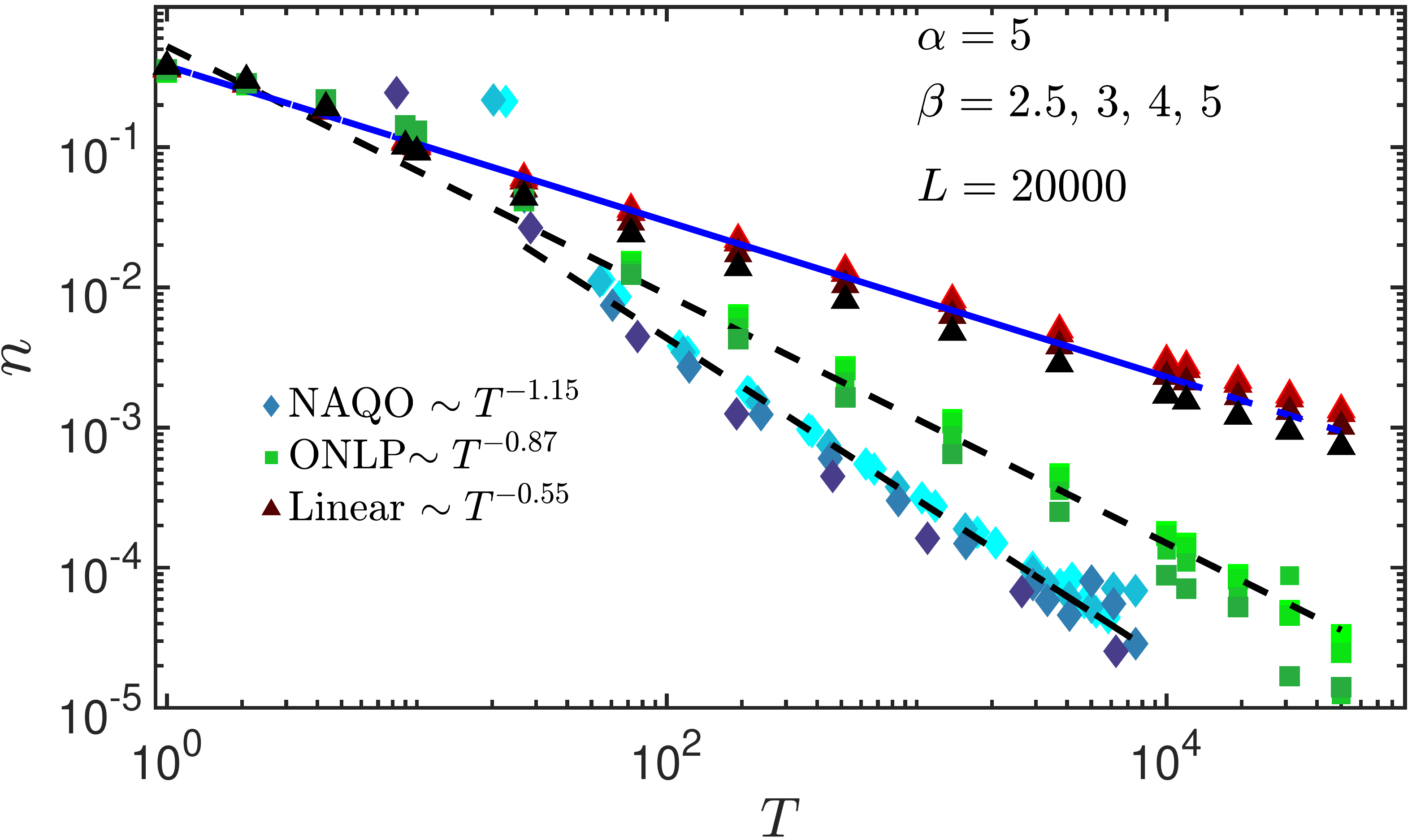}
    \caption{Universal KZ scaling of the excitation density in the short-range hopping and pairing sector of LRKMs, for $\alpha=5$ and $L=20000$. The schedule $g(t)$ obtained with the NAQO procedure provides a faster decay than the ONLP and linear schedules. One can see that the $\beta\rightarrow\infty$ limit is effectively reached around $\beta=4$.}
    \label{fig:beta>2alpha=5}
\end{figure}

\begin{figure}   
    \includegraphics[width=0.48\textwidth]{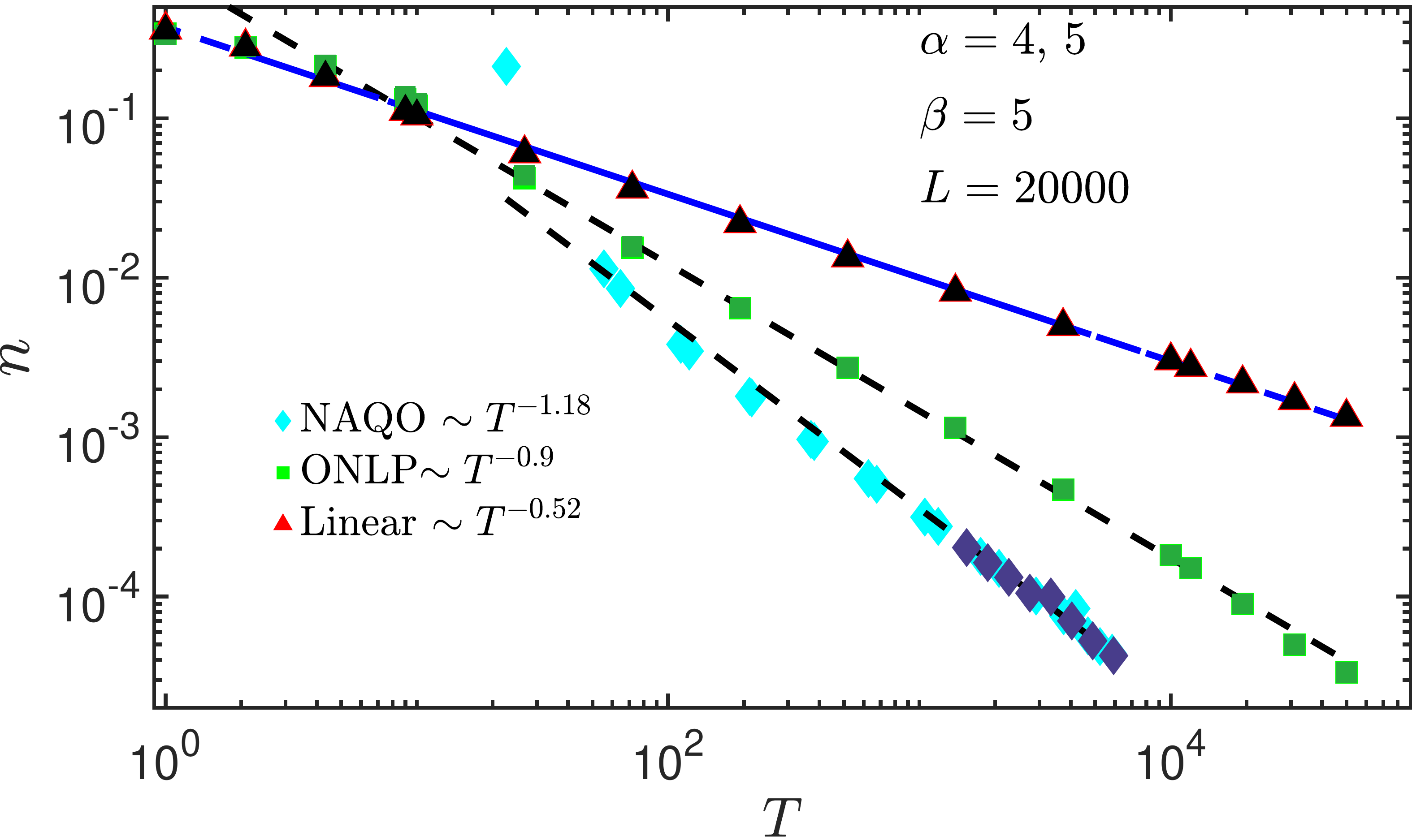}
    \caption{Universal KZ scaling of the excitation density in the short-range hopping and pairing sector of the LRKMs for $\beta=5$ and $L=20000$. This plot shows that the curves $n(T)$ are independent of $\alpha$ in the regime considered. The NAQO approach provides similar improvement over the linear and the ONLP methods.}
    \label{fig:beta>2alpha4,5}
\end{figure}

Despite bearing of paramount importance in recent developments of quantum devices and quantum technology~\cite{QMcotnrol_computation_anyons_RevMod2008, Cirac_MZM_2013, Albrecht2016MZM_edgestates}, no methods have so far been proposed to speed up the adiabatic ground state preparation of MZMs in topological superconductors. To this end, we extend the NAQO framework in Sec.~\ref{sec:NAQO} to the LRK models.

\begin{figure*}    
    \includegraphics[width=\textwidth]{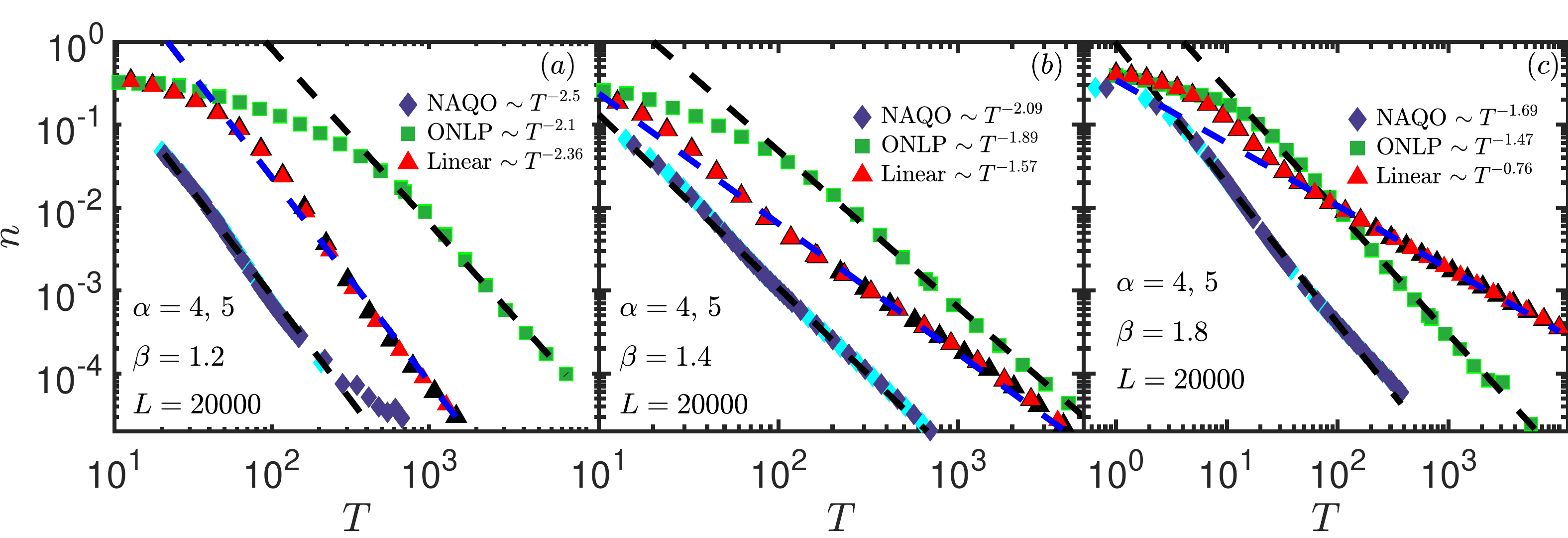}
    \caption{Scaling of the density of excitations in the KZ scaling regime of the LRKM, with long-range hopping ($\beta<2$) and short-range pairing interactions ($\alpha>3$). Three different values of $\beta$ are used, as specified in the panels. The performance of the linear schedule improves as $\beta$ decreases when compared to the ONLP. However, the NAQO schedule always exhibits the fastest decay. Results are manifestly independent of $\alpha$. }
    \label{fig:beta<2alpha>3}
\end{figure*}

For comparison, the linear ramp and the ONLP schedules are considered. The latter is expected to improve the suppression of defects similarly to the case of the TFIM~\cite{Barankov08}, as it relies on the equilibrium properties of the system~\cite{Polkovnikov2005Universal}. In addition, the ONLP takes the same form as in the TFIM, independently of the long-range exponents, as it depends only on the gap-closing exponent $z\nu$:
\begin{equation}
    g(t)=2\left(1-\left\lvert\frac{t}{T/2}\right\rvert^r\mathrm{sgn}(t)\right),
\end{equation}
where the exponent is given by
\begin{equation}\label{eq: LRK_ONLP}
    r=\frac{1}{z\nu}\ln\left[\frac{\ln(TC)}{TC}\right],
    \qquad C\approx 14.6.
\end{equation}

Due to the highly complicated $k$ dependence of the energy gaps in each momentum mode, only the leading order behavior is considered within the NAQO procedure. This still provides a considerable improvement since it takes into account all the ``dangerous'' avoided level crossings with small gaps. The leading order behavior of the energy gaps and the positions of the first avoided crossing sets the initial condition $g(0)=g_c(k_0)$. In the short-range hopping and pairing regime ($\alpha>3$, $\beta>2$), the differential equation for the optimal schedule reads
\begin{align}
    \label{eq:LRK_diffeq_1}
    \ddot g &=\frac{\zeta(\alpha-2)\zeta^2(\beta)}{4\pi\zeta(\alpha)\zeta^2(\beta-1)}\frac{\dot g^3}{(2-g)^2}-\frac{\dot g^2}{2-g}, \\
    g(0) &=2-\frac{\zeta(\alpha-2)}{\zeta(\alpha)}k^2_0,
\end{align}
where the initial conditions were set by the position of the lowest momentum state $k_0=\pi/L$.
Details of the calculations are presented in App.~\ref{app:LRK}.

Note that Eq.~\eqref{eq:LRK_diffeq_1} resembles the compact result of the TFIM. In particular, the lowest $k$ modes with vanishing gaps are accounted for via the first term on the r.h.s. with the same power of the first derivative, $\dot g^3$, and the same relation with the minimum gap for low momentum modes. Also, the second term in the r.h.s.,\ taking into account the possible transitions at higher momenta, remains intact, $\sim \dot{g}^2/(2-g)$. In both terms, the only difference arises in the approximation of $4-g^2\approx 4\,(2-g)$, owing to the small $k$ approximation. However, this does not alter the overall effects of the small gaps. 

These features are not altered much in the long-range hopping sector ($\alpha>3$, $\beta<2$), where
\begin{align}\label{eq:LRK_diffeq_beta<2alpha>3}
    \ddot g&=\frac{(2\beta-3)\zeta(\beta)^2\zeta(\alpha-2)^{\beta-1}}{4\pi\cos^2(\beta\pi/2)\Gamma(1-\beta)^2\zeta(\alpha)^{\beta-1}}\frac{\dot g^3}{(2-g)^\beta} \nonumber \\
    & \quad-(\beta-1)\frac{\dot g^2}{2-g}, \\
    g(0)&=2-\frac{\zeta(\alpha-2)}{\zeta(\alpha)}k^2_0.\quad
\end{align}
Solely, the gap dependence and the corresponding denominator in the first term of the r.h.s.\ is altered in agreement with the low momentum gap expansion in Eq.~\eqref{eq:Gap_relations}. Additionally, both terms on the r.h.s.\ also acquire $\beta$-dependent prefactors $\sim\mathcal O(1)$ that do not significantly alter the overall structure of the optimal schedule.

\begin{figure*}    
        \includegraphics[width=\textwidth]{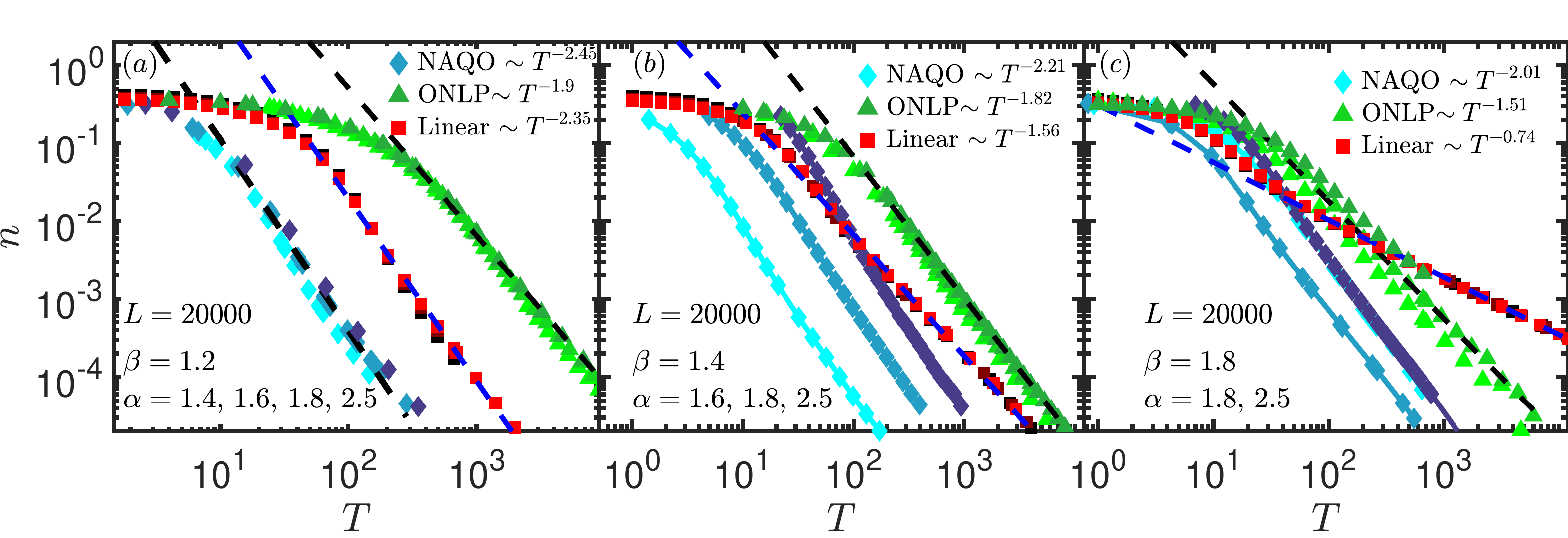}       
    \caption{Scaling of the excitation density in the KZ scaling regime of the LRKM, with short-range hopping and pairing interactions ($\beta\leq\mathrm{min}\{2,\alpha\},\,\alpha<3$). While the ONLP and linear schedules exhibit similar features to $\alpha>3$, Fig.~\ref{fig:beta<2alpha>3}, the NAQO schedule depends on the $\alpha$ exponent approximately up to a constant factor. In particular, the defect density decreases with increasing $\alpha$.}
    \label{fig:beta<alpha<3}
\end{figure*}

In the long-range pairing sector, $\alpha<3$, with the hopping interaction bounded as $\beta<\mathrm{min}\left\{\alpha,2\right\}$, the optimal control differential equation reads
\begin{widetext}
\begin{equation}\label{eq:LRK_diffeq_beta<alpha<3}
\begin{split}
&\ddot g=\frac{2\beta-\alpha}{\alpha-1}
    \frac{4\zeta(\beta)^2}{\pi\cos^2(\beta\pi/2)\Gamma(1-\beta)^2}\left(-\frac{\zeta(\alpha)}{2\Gamma(1-\alpha)\sin(\alpha\pi/2)}\right)^{\frac{2(1-\beta)}{\alpha-1}}
    \frac{\dot g^3}{(2-g)^{\frac{2\beta+\alpha-3}{\alpha-1}}}-2\frac{\beta-1}{\alpha-1}\frac{\dot g^2}{2-g},\\
    &g(0) = 2+2\sin(\alpha\pi/2)\frac{\Gamma(1-\alpha)}{\zeta(\alpha)}k^{\alpha-1}_0.
    \end{split}
    \end{equation}
\end{widetext}
In this regime, the behavior of the low momentum modes slightly changes, and a more pronounced role goes to their suppression, as entailed by $\dot g^3 /(2-g)^{(2\beta+\alpha-3)/(\alpha-1)}$. The second term, however, only acquires an $\alpha$ and $\beta$ dependent prefactor, accounting in the same way for the higher momentum excitations. This feature is attributed to the additional $\alpha$-dependent exponent in the corresponding differential equation. Despite these additional features, the resulting NAQO schedule provides the fastest decay of the defects for all $\alpha<3$.

\begin{figure*}   
        \includegraphics[width=\textwidth]{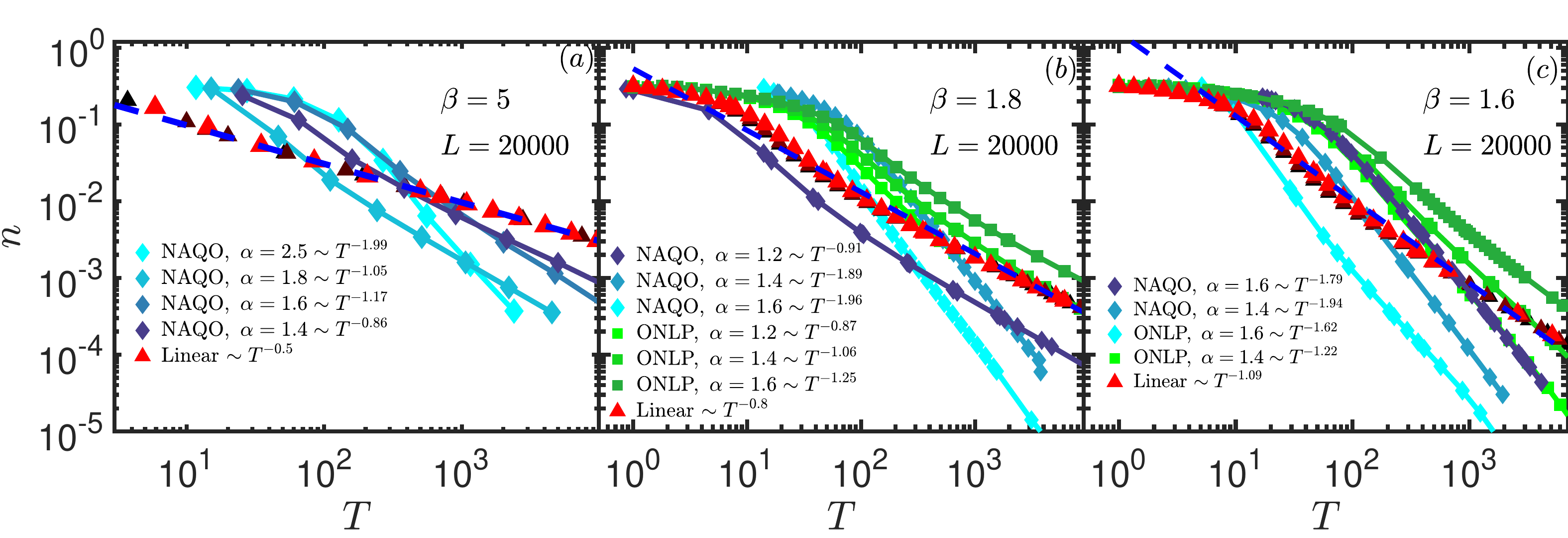}  
    \caption{Scaling of the density of excitations in the dynamical power-law-scaling regime of the LRKM for both the short-range ($\beta=5$, panel (a)) and long-range ($\beta=1.8,\,1.6$ respectively in panels (b) and (c)) hopping interactions. Linear schedules lead to results independent of $\alpha$. The NAQO schedule provides an improvement compared to both the linear and the ONLP schedules. Upon decreasing $\alpha$, the decay of the densities gets slower, and all curves converge towards that of the linear schedule.}
    \label{fig:beta>alpha<3}
\end{figure*}

The numerically obtained results are in good agreement with those presented in Sec.~\ref{sec:NAQO} for the TFIM. Within the long-range hopping and pairing sector, the same level of improvement is found. As observed in Figs.~\ref{fig:beta>2alpha=5} and \ref{fig:beta>2alpha4,5}, the three compared procedures are roughly independent of the $\alpha$ exponent, and the NAQO procedure provides a faster decay of the defect density. For values of $\beta$ in the long-range sector, the same $\alpha$ independence is observed for $\alpha>3$; see Fig.~\ref{fig:beta<2alpha>3}. Remarkably, as $\beta$ decreases, the difference between the linear, ONLP, and NAQO results disappears. In this limit, the NAQO scheme still provides an improvement in the KZ scaling with a finite constant factor, as observed in Figs.~\figpanel{fig:beta<2alpha>3}{a,b} and \figpanel{fig:beta<alpha<3}{a,b}. This feature immediately follows from the structure of Eq.~\eqref{eq:LRK_diffeq_beta<2alpha>3}, as the term on the r.h.s.\ disappears when $\beta\rightarrow1$. It is noteworthy that both in Eq.~\eqref{eq:LRK_diffeq_beta<2alpha>3} and Eq.~\eqref{eq:LRK_diffeq_beta<alpha<3} the resulting differential equation reads
\begin{equation}
    \ddot g\sim \frac{\dot g^3}{2-g}.
\end{equation}
Near the critical point, the solution becomes linear since the diverging denominator cancels the r.h.s.,\ implying a vanishing second derivative. For larger momenta, the solution that satisfies the boundary conditions is
\begin{equation}
    g(t)=2+\frac{2-t/T}{W((t/T-2) L^2 e^{-L^2/\pi^2}/\pi^2)},    
\end{equation}
where again $W$ is the Lambert function, cf.\ Eq.~\eqref{eq:Lambert}. At small $t$ values, the denominator can be neglected, while for intermediate values, the shape stays close to the linear one as well.
Note that this limiting differential equation is of completely opposite nature to that in the eQAB,  Eq.~\eqref{eq:eQAB_diff_eq}, which only involves the $\dot g^2/(2-g)$ term.

Finally, in the case of Eq.~\eqref{eq:LRK_diffeq_beta<alpha<3}, the $\alpha$ independence is only observed for the linear and ONLP schedules, while increasing $\alpha$ shifts the LZ-optimized schedules by a constant factor, as observed in \figpanel{fig:beta<alpha<3}{b,c}.

\subsection{Optimal control in the dynamical power-law-scaling regime}
\label{sec:LRK_dyn_univ_regime}

Within the range of exponents $\alpha<\beta$, $\alpha<3$, the defect density exhibits a new dynamical power-law scaling that does not match the predictions of the KZ mechanism. As pointed out in Ref.~\cite{Defenu_LRK}, the power law is determined solely by the $\beta$ exponent for linear ramps:
\begin{equation}
    \label{eq:linearbeta<alpha<3}
    n \sim T^{-1/(2\beta-2)}.
\end{equation}
The equilibrium critical exponents $z=\alpha-1$ and $\nu=1/(\alpha-1)$, on the other hand, would suggest the scaling $n \sim T^{-1/(2\alpha-2)}$. For this reason, the ONLP schedule loses its validity, being it reliant on the equilibrium properties. To demonstrate the supremacy of the NAQO even in this regime, one can still stick to the comparison with respect to the linear and KZ-based ONLP schedules.

\begin{figure*}   
        \includegraphics[width=\textwidth]{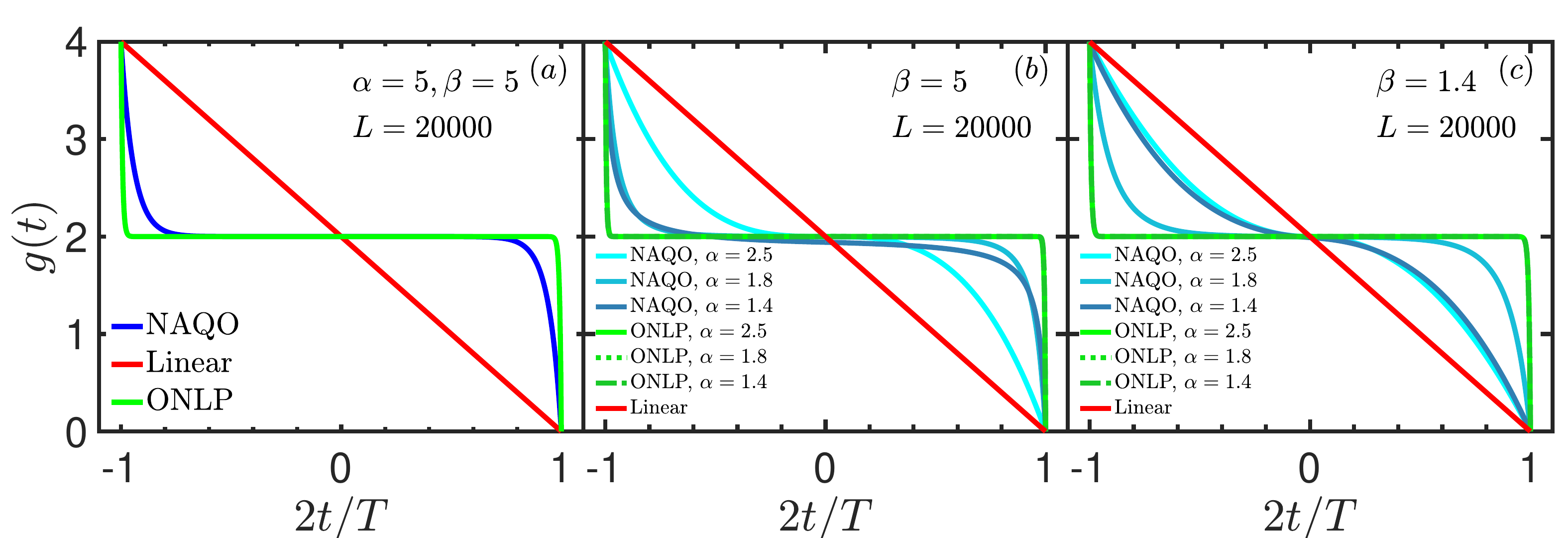}  
    \caption{Optimal NAQO schedules for $(a)$ the KZ scaling regime ($\beta=5,\,\alpha=5$), $(b)$ the short-range hopping regime (($\beta=5,\,\alpha=1.4,\,1.8,\,2.5$), and $(c)$ the long-range case ($\beta=1.4,\,\alpha=1.6,\,\alpha=1.8,\,\alpha=2.5$). The evolution times $T$ is taken near $T_\mathrm{ons}$. The schedules are compared with the ONLP and linear ones. While these latter do not depend on $\alpha$ up to good precision, the NAQO curves accurately take into account the different pairing effects.}
    \label{fig:LRK_protocols}
\end{figure*}

The dynamical universality regime can also be split into the long-range, $\beta<2$, and short-range, $\beta>2$, hopping sectors. The optimal control differential equation for the latter case reads (for details, see App.~\ref{app:LRK})
\begin{multline}
    \ddot g=\frac{4-\alpha}{\alpha-1} 
    \frac{4\zeta^2(\beta) \dot g^3}{\pi\zeta^2(\beta-1) (2-g)^{\frac{\alpha+1}{\alpha-1}}} \\
    \times \left(\frac{-\zeta(\alpha)}{2\Gamma(1-\alpha)\sin(\alpha\pi/2)}\right)^{\frac{2}{1-\alpha}} -\frac{2}{\alpha-1}\frac{\dot g^2}{2-g},
\end{multline}
with initial condition
\begin{equation}    
    g(0)  =2+2\sin(\alpha\pi/2)\frac{\Gamma(1-\alpha)}{\zeta(\alpha)}k^{\alpha-1}_0,
\end{equation}
where notably the low-momentum term on the r.h.s., $\dot g^3/(1-g)^{\frac{\alpha+1}{\alpha-1}}$, only depends now on $\alpha$. In this regime, both ONLP and NAQO provide a faster decay of $n(T)$ as $\alpha$ increases, while the linear schedule results remain unchanged according to Eq.~\eqref{eq:linearbeta<alpha<3}, as predicted in Ref.~\cite{Defenu_LRK}. 

The ONLP schedules are independent of $\alpha$ by construction. However, the corresponding excitation densities exhibit a slight dependence on it; this is attributed to the different spectral structures that are traversed by the same schedule. 

The NAQO schedules depend strongly on $\alpha$, perfectly accounting for the dynamical properties governed by the long-range pairing interactions. In particular, the NAQO density of defects falls off faster for increasing $\alpha$, and it converges towards the linear schedule as $\alpha\rightarrow1$, as shown in Fig.~\figpanel{fig:beta>alpha<3}{a}. Furthermore, in agreement with the previous results for all $\alpha$ values in this regime, the NAQO procedure provides considerable improvement compared to both the linear and ONLP results. The power-law scaling curves were plotted for $\beta=5$, similar to the long-range hopping sector inside the KZ scaling regime. The defect density becomes independent of the hopping interaction around $\beta\approx4$.

In the more interesting long-range hopping sector, $\alpha<\beta<2$, the defect density depends on both exponents in the ONLP and NAQO procedures, while it remains invariant with respect to $\alpha$ in the case of the linear ramp. Similar to the short-range hopping regime, $\beta>2$, the performance of the ONLP is not analytically controlled as it cannot capture the dominating dynamical effects of the hopping interactions. However, the NAQO approach correctly accounts for both the pairing and hopping effects in this regime as well. As demonstrated in Figs.~\figpanel{fig:beta>alpha<3}{b} and \figpanel{fig:beta>alpha<3}{c}, the fastest density decay is given by the NAQO protocols, while the ONLP exhibits slightly better performance than the linear ramp. However, both curves converge towards the linear one as $\alpha$ approaches one.

In Fig.~\ref{fig:LRK_protocols}, the obtained NAQO and ONLP schedules are displayed for different values of $T_\mathrm{ons}$. In the TFIM-like regime, all curves are independent of both long-range exponents. The same applies to the ONLP schedule in both the long-range hopping and pairing regime, $\beta<\alpha<3$, and short-range hopping and long-range pairing regime, $\beta=5$ and $\alpha<3$, within the dynamical power-law-scaling regime, in agreement with Eq.~\eqref{eq: LRK_ONLP}. The only weak dependence stems from the slightly different $T_\mathrm{ons}$ values. The NAQO schedules perfectly account for the additional characteristics of the pairing interactions, as completely distinct curves are observed as $\alpha$ changes.

\section{Demonstration on a non-integrable disordered many-body system}
\label{sec:non-integrable}

In this last section, the efficiency of the NAQO framework is probed on generic non-integrable spin systems that do not admit a free-fermionic representation. We consider the Hamiltonian~\cite{Pollmann_MBL} 
\begin{equation}\label{eq:H_Non_int}
	\hat{H}(t) = 
        - \sum_{j=1}^L \left[ \left(J+\delta J_j\right)\,\hat{\sigma}^z_j \hat{\sigma}^z_{j+1}-J' \hat\sigma^z_j\hat\sigma^z_{j+2} + g(t) \hat{\sigma}_j^x \right]\,,
\end{equation}
where the disordered couplings $\delta J_j$ are extracted from the uniform distribution over $[-\delta J,\,\delta J]$. The next-nearest-neighbor $J'$ coupling has been introduced to break integrability and to generate some frustration, in analogy to the NP-hard spin-glass problems usually considered in quantum annealing~\cite{Lucas_Np_hard_review}. The disorder $\delta J$, no matter how small, is a relevant perturbation to the clean model according to the Harris criterion~\cite{Harris1974Effect,Cardy1996Scaling}, and changes the universality class to that of the infinite-disorder fixed point introduced in Refs.~\cite{Fisher1992Random,Fisher1995Critical}. As a consequence, the annealing dynamics of disordered Ising models undergo a significant slow-down or, equivalently, the number of defects decreases only as a logarithm of the final time $T$, rather than a power law, for large enough system sizes~\cite{Dziarmaga2006Dynamics,PYoung2008,PYoung2010}. For the numerical simulations, $\delta J=0.1,\,J^\prime=0.4\,J$ were chosen, while the energy unit is set to $J=1$ like in the previous sections.

The model in Eq.~\eqref{eq:H_Non_int} does not admit a separation into independent momentum modes, but instead, the Hilbert space is composed of only two $2^{L-1}$-dimensional sectors, according to the parity $\prod_i \hat{\sigma}_i^x$. We concentrate on the even parity sector, which is the one dynamically accessible from the paramagnetic ground state. 

\begin{figure}
    \includegraphics[width=.48\textwidth]{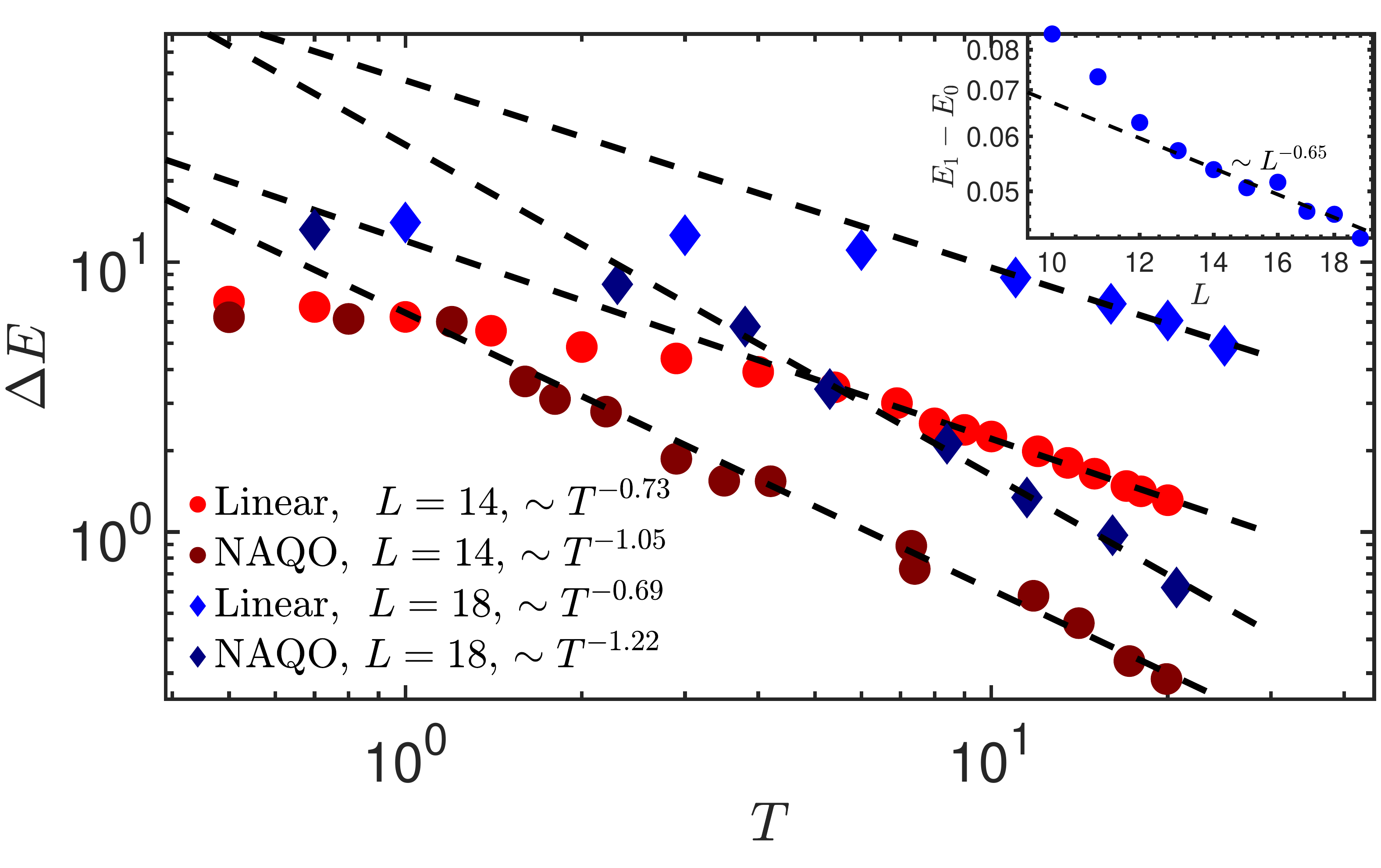} 
    \caption{Excess energy density as a function of $T$ for system sizes $L=14$ and $L=18$ for the linear schedule and the NAQO with the numerically extracted critical point and power-law scaling of the gap. The NAQO results show a clear advantage with a similar improvement in the power-law as in the LRKM case. The inset shows the decreasing power law of the gap as a function of $L$.}
    \label{fig:Non_Int}
\end{figure}

As in the integrable cases, excitations are generated mainly via the LZ transitions at multiple avoided crossings. Although these transitions are no longer independent of each other, and their analytical treatment is not possible, some robust features are still present. In particular, avoided crossings of higher excited states appear with gradually increasing gaps and at smaller values of $g(t)$, motivating the framework of the NAQO. As a stepping stone towards a more precise numerical treatment of these avoided crossings, we employ an ansatz approach in terms of the general LRKM control equation. Even though the critical exponents are not accessible analytically, within this ansatz approach, we identify the $\beta$ parameter by the $L$ dependence of the ground state gap.
For the numerically achievable system sizes $L\lesssim20$, the ground-state energy gap approximately decreases as $\sim L^{-0.65}$ (see inset of Fig.~\ref{fig:Non_Int}).
The general LRKM equation in long-range hopping and short-range pairing regime $\beta<2,\,\alpha>3$ (see Sec.~\ref{sec:LRK}) reads
\begin{align}
    \ddot g&=\frac{(2\beta-3)\zeta(\beta)^2\zeta(\alpha-2)^{\beta-1}}{4\pi\cos^2(\beta\pi/2)\Gamma(1-\beta)^2\zeta(\alpha)^{\beta-1}}\frac{\dot g^3}{(2-g)^\beta} \nonumber \\
    & \quad-(\beta-1)\frac{\dot g^2}{2-g}.
\end{align}
In the first term on the r.h.s., the $L$ dependence of the gap at the first avoided level crossing is captured by $\beta$. The second term accounts for the suppression of LZ transitions to higher excited states. For the numerically reachable system sizes, $L\lesssim20$, the effective system size dependence of the ground state gap sets the exponent based on the dispersion relation of the LRMs, $\beta=1.65$, while the second term on the r.h.s. provides a good approximation for the higher excited states as it depends on $\beta$ only through a multiplicative factor. Additionally, the limit $\alpha\rightarrow\infty$ was taken to illustrate the use of this ansatz in the simplest possible form. The initial condition of the corresponding optimal differential equation is identified with the position of the gap to the ground state, also found numerically. Thus, the corresponding optimal control equation reads
\begin{align}
    \ddot g=\frac{(2\beta-3)\zeta(\beta)^2}{4\pi\cos^2(\beta\pi/2)\Gamma(1-\beta)^2}\frac{\dot g^3}{(2-g)^\beta}-(\beta-1)\frac{\dot g^2}{2-g}\,,
\end{align}
with the numerically extracted $\beta\approx1.65$. Note that the equation above sets the critical point close to $2$, but this issue is readily solved by rescaling the schedule with the position of the gap.

We remark that the numerical extraction of the $\beta$ effective power and the position of the minimal gap involves a much smaller numerical effort than performing the time evolution.
In our case, the scaling of the ground state gap and its position was extracted by sparse matrix diagonalization methods. The time evolution was performed by a Chebyshev polynomial representation operator developed recently in Ref.~\cite{BADER_SkewHermitian_2023}.

The results are shown in Fig.~\ref{fig:Non_Int} for $L=14$ nd $L=18$. The present method provides a tangible speed-up compared to earlier path optimization methods~\cite{Zeng_PathOptimization_2016,Cote_2023_FreeFermionOPtContr,Quiroz2019_GroverNum}, which rely exclusively on numerical optimization and typically require also the use of intermediate control Hamiltonians. In these studies, both the computational resources and the achievable efficiency are much larger than the NAQO method either semi-analytical in integrable models or assisted by additional numerical techniques in non-integrable ones.

\section{Conclusions and perspectives}
\label{sec:conclusions}

Minimizing topological defect formation when crossing a QPT is of paramount importance in quantum technological applications, especially in the high-fidelity preparation of novel phases of matter.  
Oftentimes, implementing additional control fields to speed up the passage across the critical point is not possible or convenient experimentally, and a feasible strategy is provided by tailoring the schedule of a single control field at disposal.

In this work, we showed that optimal control strategies relying on the adiabatic condition fail to drive free-fermionic models, such as the transverse-field Ising model (TFIM) and the long-range Kitaev model (LRKM), in a timescale shorter than the adiabatic one. By expressing the models into independent momentum modes, it was shown analytically that the avoided crossings in higher momentum modes generate an extensive number of excitations, even when the adiabatic condition is satisfied locally. The mechanism was tested via the Kibble-Zurek analysis of the density of excitations and via the timescale for the onset of adiabaticity.

It was found that, in the TFIM, the Cerf-Roland local adiabatic driving (LAD) based on the ground state alone performs worse than the simple linear passage. Further, the quantum adiabatic brachistochrone (QAB) provides only a slight improvement. To account for the higher-momentum transitions, we also introduced an extended QAB (eQAB), which, however, was only able to improve the onset of adiabaticity timescale by a constant factor.

Our central result is the introduction of an analytical Non-Adiabatic Quantum Optimization  (NAQO) framework based on the optimization of LZ transition probabilities in each momentum mode. The only numerics required to apply this approach consists of solving an ordinary differential equation, thus it can be implemented without the requirement of heavy optimization algorithms. We showed that the NAQO scheme provides a sizeable speed-up if compared to the optimization strategies based on the adiabatic condition, the KZ mechanism (like the ONLP), and numerical optimization, which is also restricted to small system sizes.

The general framework was also applied to the LRKM with power-law-decaying hopping and pairing interactions. Our study provides efficient speed-up schedules for the crossing of its QPT, while the other methods fail in some regimes of the long-range interaction strengths. Specifically, the LRKM hosts four different regimes, according to the exponents $\alpha$ and $\beta$, regulating the spatial decay of hopping and pairing. In the TFIM-like regime with weak pairing and hopping, the results for the LRKM are comparable to those for the TFIM, as expected. Then, as long as the Kibble-Zurek mechanism holds, the NAQO is found to perform considerably better than the ONLP and the linear schedules. However, the defect densities obtained with the three schedules become comparable upon increasing the hopping range. In the more interesting dynamical universality regime, where pairing interactions dominate, the ONLP loses its validity as it relies on the equilibrium properties of the system, which no longer describe the density of excitations. However, even in this regime the NAQO successfully captures the dynamical properties, providing again the best decay of defects as a function of the evolution time. 

Our work provides a comparative study of analytical optimization methods for analytically treatable models, thus setting the base for further investigations in more complicated cases. Contrary to the optimization of a single qubit (or few qubits), in the case of the TFIM and LRKM, the optimization scheme needs to take into account the presence of several excitation channels, whose avoided crossings take place at different values of the control parameter, and are of different shape. For this reason, the use of the LZ formula in each momentum mode, instead of the adiabatic condition, leads to the best faster-than-adiabatic control schedules. 

A challenging and promising prospect is to apply  NAQO and the LZ formula to non-integrable models and in more general quantum annealing contexts. While we have demonstrated for the specific example of Sec.~\ref{sec:non-integrable} that NAQO does provide some speed-up at small system sizes, it is an open question whether NAQO is able to speed up the annealing also when, at larger sizes, the scaling of the gap becomes exponential in the system size~\cite{Dziarmaga2005Dynamics}. The same exponential scaling is at the base of the difficulty of applying quantum annealing to NP-hard problems~\cite{PYoung2008,PYoung2010}. In this context, optimal control is still largely uncharted, and the few existing studies point to the presence of interesting phases and controllability transitions~\cite{Day2019Glassy,Ge2022Optimization,Dalgaard2022Predicting,Beato2024Theory}. Another interesting direction would be to use the analytically-optimized NAQO schedules as a starting point for further numerical optimization in the spirit of Refs.~\cite{Wu2015Optimal,Hernandez2021Optimal}.

\acknowledgments

We thank Pranav Chandarana and Xi Chen for the insightful discussions. The numerical simulations presented in this work were carried out using the HPC facilities of the University of Luxembourg. This project was supported by the Luxembourg National Research Fund (FNR Grant Nos.\ 17132054 and 16434093). It has also received funding from the QuantERA II Joint Programme and co-funding from the European Union’s Horizon 2020 research and innovation programme.

\appendix

\section{Differential equation for the quantum adiabatic brachistochrone}
\label{app:sec:QAB}

This appendix shows how to integrate the differential equation~\eqref{eq:EL_QAB_TFIM}, that is reproduced here for convenience:
\begin{equation}
    \ddot{g} = \frac{2 \dot{g}^2(g-\cos k_0)}{1+g^2-2g\cos k_0}.
\end{equation}
Being the equation autonomous, one can define $p(g) \equiv \dot{g}$, so that $\ddot{g} = p'(g) p(g)$ and
\begin{equation}
    \frac{p'}{p} = \frac{2(g-\cos k_0)}{1+g^2-2g\cos k_0}.
\end{equation}
This equation can be integrated directly, leading to
\begin{equation}
    p(g) = p_0 (1+g^2-2g\cos k_0).
\end{equation}
Now, recalling that $p(g) = \dot{g}$, one can integrate again, finding
\begin{align}
    p_0(t-t_0) &= \int \frac{\mathrm{d}g}{1+g^2-2g\cos k_0} \nonumber\\
    &= \frac{1}{2i\sin k_0} \int \mathrm{d}g \left[ \frac{1}{g-e^{ik_0}}-\frac{1}{g-e^{-ik_0}} \right] \nonumber\\
    &= \frac{1}{2i\sin k_0} \log \frac{g-e^{ik_0}}{g-e^{-ik_0}}.
\end{align}
Solving the equation above for $g$, one finds
\begin{align}
    g(t) &= \frac{e^{ik_0}-e^{-ik_0-2ip_0 (t-t_0)\sin k_0 }}{1-e^{-2ip_0 (t-t_0)\sin k_0 }} \\
    &= 2\frac{\sin[k_0 + \varepsilon(t-t_0)] \sin[\varepsilon(t-t_0)]}{1-\cos[2\varepsilon(t-t_0)]},
\end{align}
where $\varepsilon = -p_0 \sin k_0$. Finally, due to the freedom in setting the zero of the time axis, it is convenient to shift $t_0 \to t_0 - \pi/2\varepsilon$ in order to have a schedule (almost) symmetric around $t=0$:
\begin{equation}
    g_\mathrm{QAB}(t) = \cos k_0 - \sin k_0 \tan [\varepsilon(t-t_0)],
\end{equation}
which is Eq.~\eqref{eq:g_QAB} in the main text.

\section{Differential equation for the extended quantum adiabatic brachistochrone}
\label{app:sec:eQAB}

The Euer-Lagrange equation for the extended quantum brachistochrone (eQAB) action is found from
\begin{equation}
    \fd{S_\mathrm{eQAB}}{g} = -\frac{10\dot g^2\left[g(g^2-1)+(1+2g)k_0\right]}{\left[(g^2-1)^2+4g(1+g^2)k_0\right]^{5/2}}
\end{equation}
and
\begin{multline}
    \der{}{t} \fd{S_\mathrm{eQAB}}{\dot{g}} = \frac{2\ddot g}{\left((g^2-1)^2+4g(1+g^2)k_0\right)^{3/2}} \\ 
    -\frac{20\dot g^2\left[g(g^2-1)+(1+2g)k_0\right]}{\left[(g^2-1)^2+4g(1+g^2)k_0\right]^{5/2}}.
\end{multline}
Putting the two together, one finds
\begin{align}
    \ddot g &=\frac{5\dot g^2\left[g(g^2-1)+(1+2g)k_0\right]}{(g^2-1)^2+4g(1+g^2)k_0} \\
    &\approx \frac{5\dot g^2\left[g(g^2-1)\right]}{(g^2-1)^2+8k_0}\,,
\end{align}
where in the last step, $k_0$ was dropped from the numerator and $4g(1+g^2)\approx 8$ was used in the denominator.

\section{Differential equation for the NAQO approach}
\label{app:sec:NAQO}

The Euler-Lagrange equation is found by computing
\begin{equation}
    \fd{S_\mathrm{NAQO}}{g} = \left[-\frac{g\dot g}{(1-g^2)^{3/2}}+4\pi\frac{g}{\sqrt{1-g^2}}\right]e^{-2\pi\frac{1-g^2}{\dot g}}
\end{equation}
and
\begin{equation}
    \der{}{t} \fd{S_\mathrm{NAQO}}{\dot g} = \der{}{t} \left[ \left( \frac{1}{\sqrt{1-g^2}}+2\pi\frac{\sqrt{1-g^2}}{\dot g}\right) e^{-2\pi\frac{1-g^2}{\dot g}} \right].
\end{equation}
After equating the two, a lengthy but straightforward computation leads to Eq.~\eqref{eq:LZ_opt_diffeq}.

\section{NAQO action in the long-range Kitaev model}
\label{app:LRK}

In this appendix, we detail the derivation of the NAQO differential equation for the LRKM. As shown in Ref.~\cite{Defenu_LRK} and summarized in Sec.~\ref{sec:LRK}, the LRK Hamiltonian with power-law long-range hopping and pairing interactions has single-particle energy levels given by Eq.~\eqref{eq:LRK_SP_energies}. These levels form avoided level crossing at the local minima of the difference of TLS energies. The avoided crossing positions can be obtained exactly in terms of $j_\alpha(k)$ and $d_\beta(k)$ as
\begin{align}
    &\partial_g\sqrt{\left(g(t)/2-j_\alpha(k)\right)^2+d^2_\beta(k)}=0\\
    &\implies \quad \left(g(t)/2-j_\alpha(k)\right)\partial_g\left(g(t)/2-j_\alpha(k)\right)=0 \\
    &\implies \quad g_c(k)=2\sum_rj_{r,\alpha}\cos(kr)\,.
\end{align}
This rather complicated relation simplifies in the small-$k$ limit, which still captures all the essential features of low-energy excitations governing the non-adiabatic regime for $\alpha>1$:
\begin{align}
    g_c(k)\approx
    \begin{cases}
        2+2\sin(\alpha\pi/2)\frac{\Gamma(1-\alpha)}{\zeta(\alpha)}k^{\alpha-1} &\alpha<3\\ 
        g_c(k)\approx2-\frac{\zeta(\alpha-2)}{\zeta(\alpha)}k^2                &\alpha>3.
    \end{cases}
\end{align}
The LZ excitation probabilities are governed only by the $\beta$ exponent:
\begin{equation}
    p_k\approx e^{-2\pi d^2_\beta(k) / \dot g_c(k)}\,,
\end{equation}
where $\dot g_c(k)$ denotes the time-derivative of the schedule at the local gap minimum. Thus, the excitation density is expressed as the sum of the LZ probabilities, extended now to arbitrary schedules satisfying that $g(-T/2)=4,\,g(0)=2,\,g(T/2)=0$:
\begin{equation}
    n\approx\frac{1}{2\pi}\int\mathrm dk\,e^{-2\pi d^2_\beta(k)/\dot g_c(k)}\,.
\end{equation}
As in the TFIM, Sec~\ref{sec:NAQO}, it is convenient to switch to the time variable and transform the integral. First, express the time instance where the $k$-th avoided crossing appears. Then, performing the change of variable in the integration measure
\begin{equation}
     t=g^{-1}\left[2\sum_rj_{r,\alpha}\cos(kr)\right],
\end{equation}
leads to
\begin{align}
    \frac{\mathrm dt}{\mathrm dk} &=-\left(g^{-1}\right)^\prime\left[g(t_c)\right]\,2\sum_r j_{r,\alpha}r\sin(kr)\\
    &=\frac{2\sum_r j_{r,\alpha}r\sin(kr)}{g^\prime(g^{-1}\left[g(t_c)\right])}\\
    &=\frac{2\sum_r j_{r,\alpha}r\sin(kr)}{g^\prime(t_c)}
\end{align}
and
\begin{equation}
    \mathrm dk =\mathrm dt\,\left(\frac{\mathrm dt}{\mathrm dk}\right)^{-1}
    =\mathrm dt\,\frac{\dot g(t)}{2\sum_r \tilde j_{r,\alpha-1}\sin(kr)},
\end{equation}
where $(\dots)^\prime$ is the abbreviation of $\mathrm d /\mathrm dk$ and $t_c$ denotes the time instance when the $k$-th avoided crossing appears. Notice that the new long-range hopping interaction term has different normalization,
\begin{equation}
    \tilde j_{r,\alpha-1}=\frac{N_{\alpha-1}}{N_\alpha}j_{r,\alpha-1}.
\end{equation}
In the last step, the crossing time $t_c$ was used as the new time variable, and the derivative with respect to the control parameter was changed to a time derivative. 

The above steps lead to the final expression for the NAQO action
\begin{equation}
    S_\mathrm{NAQO}=\int\mathrm dt \, \frac{\dot g}{2\sum_r j_{r,\alpha-1}\sin(kr)}\exp\left[-2\pi \frac{d^2_\beta(k)}{\dot g}\right].
\end{equation}
Even though this is a highly complicated expression, it can still be handled analytically in the small $k$ limit in order to show how the effects of the long-range hoppings manifest. For this purpose, one can exploit the leading order expansion of the $d_\beta(k)$ terms and the modified expansions of $j_{\alpha-1}(k)$:
\begin{equation}
    d_\beta(k)\approx
    \begin{cases}
        \cos(\beta\pi/2)\frac{\Gamma(1-\beta)}{2\zeta(\beta)}k^{\beta-1} &\beta<2\\ 
        \frac{\zeta(\beta-1)}{2\zeta(\beta)}k &\beta>2.
    \end{cases}
\end{equation}
The Lagrangian of the action $S_\mathrm{NAQO}$ follows:
\begin{equation}
    \mathcal L_\mathrm{NAQO}=\frac{\dot g}{k^{\theta(\alpha-1)}}\exp\left[-C_\beta\frac{k^{2\theta(\beta)}}{\dot g}\right],
\end{equation}
where $\theta(\gamma)=(\gamma-1)\Theta(\gamma<2)+\Theta(\gamma\geq2)$ and the $k$-dependent part was written out. The structure is the same as for the TFIM for $\alpha>3,\,\beta>2$, while it gets slightly modified for $\beta<2$ and $\alpha>3$, as then the exponent of $k$ becomes less than one. In the opposite regime ($\alpha<3$, $\beta>2$), the exponent remains $k^2$, but the denominator has a power smaller than one in $k$. In the fourth case ($\alpha$, $\beta<2$), both the exponent and the denominator have a power smaller than one. The remaining $k$ dependence is expressed in terms of $g(t)$ via a first-order $k$ expansions:
\begin{align}
    \mathcal L_\mathrm{NAQO} &\approx \frac{\dot g}{k}\exp\left[-2\pi\frac{\zeta^2(\beta-1)}{\zeta^2(\beta)}\frac{k^2}{\dot g}\right] \\
    &\approx \frac{\dot g}{\sqrt{2-g}}\exp\left[-\frac{2\pi\zeta^2(\beta-1)}{\zeta^2(\beta)}\frac{\frac{\zeta(\alpha)}{\zeta(\alpha-2)}(2-g)}{\dot g}\right],
\end{align}
with boundary condition
\begin{equation}
    g(0)=2-\frac{\zeta(\alpha-2)}{\zeta(\alpha)}k^2  
\end{equation}
for $\beta>2$ and $\alpha>3$;
\begin{widetext}
\begin{align}
    \mathcal L_\mathrm{NAQO} &\approx \frac{\dot g}{k}\exp\left[-2\pi\frac{\cos^2(\beta\pi/2)\Gamma(1-\beta)^2}{\zeta(\beta)^2}\frac{k^{2(\beta-1)}}{\dot g}\right] \\
    & \approx \frac{\dot g}{\sqrt{2-g}}\exp\left[\frac{2\pi\cos^2(\beta\pi/2)\Gamma(1-\beta)^2}{\zeta(\beta)^2}\frac{\frac{\zeta(\alpha)^{\beta-1}}{\zeta(\alpha-2)^{\beta-1}}\left(2-g\right)^{\beta-1}}{\dot g}\right],
\end{align}
with boundary condition
\begin{equation}
    g(0)=2-\frac{\zeta(\alpha-2)}{\zeta(\alpha)}k^2
\end{equation}
for $\alpha>3$ and $\beta<2$;
\begin{align}
    \mathcal L_\mathrm{NAQO} &\approx \frac{\dot g}{k^{\alpha-2}}\exp\left[-2\pi\frac{\zeta^2(\beta-1)}{\zeta^2(\beta)}\frac{k^2}{\dot g}\right] \\
    &\approx \frac{\dot g}{\left(2-g\right)^{\frac{\alpha-2}{\alpha-1}}}
    \exp\left[-\frac{2\pi\zeta^2(\beta-1)}{\zeta^2(\beta)}\frac{\left(-\frac{\zeta(\alpha)}{2\Gamma(1-\alpha)\sin(\alpha\pi/2)}(g-2)\right)^{\frac{2}{\alpha-1}}}{\dot g}\right]\\
\end{align}
with boundary condition
\begin{equation}
    g(t)=2-2\sin(\alpha\pi/2)\frac{-\Gamma(1-\alpha)}{\zeta(\alpha)}k^{\alpha-1}
\end{equation}
for $\beta>2$ and $\alpha<3$;
\begin{align}
    \mathcal L_\mathrm{NAQO} &\approx \frac{\dot g}{k^{\alpha-2}}    
    \exp\left[-2\pi\frac{\cos^2(\beta\pi/2)\Gamma(1-\beta)^2}{\zeta(\beta)^2}\frac{k^{2(\beta-1)}}{\dot g}\right] \\
    &\approx \frac{\dot g}{\left(2-g\right)^{\frac{\alpha-2}{\alpha-1}}}
    \exp\left[-2\pi\frac{\cos^2(\beta\pi/2)\Gamma(1-\beta)^2}{\zeta(\beta)^2}\frac{\left(-\frac{\zeta(\alpha)}{2\Gamma(1-\alpha)\sin(\alpha\pi/2)}(g-2)\right)^{\frac{2(\beta-1)}{\alpha-1}}}{\dot g}\right]
\end{align}
with boundary condition
\begin{equation}
    g(0)=2-2\sin(\alpha\pi/2)\frac{-\Gamma(1-\alpha)}{\zeta(\alpha)}k^{\alpha-1}
\end{equation}
for $\beta<2$ and $\alpha<3$. Note that the modified prefactor in $\tilde j_{r,\alpha-1}$ does not affect these Lagrangians as they only appear as multiplicative factors.

The action $S_\mathrm{NAQO}$ is minimized by taking the variation with respect to $g(t)$. Since the NAQO Lagrangian has the general form
\begin{equation}
    \mathcal L_\mathrm{NAQO}=\frac{\dot g}{(2-g)^a}\exp\left[-b\frac{(2-g)^c}{\dot g}\right],
\end{equation}
it is convenient to compute
\begin{eqnarray}
    \partial_g\mathcal L&=&\left[a\dot g(2-g)^{-a-1}+bc(2-g)^{c-a-1}\right]\exp\left[-b\frac{(2-g)^c}{\dot g}\right],\\
    \partial_{\dot g}\mathcal L&=&\left[(2-g)^{-a}+b\frac{(2-g)^{c-a}}{\dot g}\right]\exp\left[-b\frac{(2-g)^c}{\dot g}\right]  ,  
\end{eqnarray}
and
\begin{multline}
    \frac{\mathrm d}{\mathrm dt}\partial_{\dot g}\mathcal L=\Big[
    a\dot g(2-g)^{-a-1}-b\frac{(2-g)^{c-a}}{\dot g^2}\ddot g-b(c-a)(2-g)^{c-a-1}\\
    +b^2\frac{(2-g)^{2c-a}}{\dot g^3}\ddot g+cb^2\frac{(2-g)^{2c-a-1}}{\dot g}+b\frac{(2-g)^{c-a}}{\dot g^2}\ddot g+bc(2-g)^{c-a-1}
    \Big]
    \exp\left[-b\frac{(2-g)^c}{\dot g}\right].
\end{multline}
The Euler-Lagrange equation follows,
\begin{equation}
    -b(c-a)(2-g)^{c-a-1}+b^2\frac{(2-g)^{2c-a}}{\dot g^3}\ddot g+cb^2\frac{(2-g)^{2c-a-1}}{\dot g}=0,
\end{equation}
and simplifying,
\begin{equation}
    \ddot g=\frac{c-a}{b}\frac{\dot g^3}{(2-g)^{c+1}}-c\frac{\dot g^2}{2-g}.
\end{equation}
We next specify this equation in each of the four cases:
\begin{align}
    \ddot g &=\frac{1\zeta(\alpha-2)\zeta^2(\beta)}{4\pi\zeta(\alpha)\zeta^2(\beta-1)}\frac{\dot g^3}{(2-g)^2}-\frac{\dot g^2}{2-g},\\
    g(0) &=2-\frac{\zeta(\alpha-2)}{\zeta(\alpha)}k^2_0,
\end{align}
for $\beta>2$ and $\alpha>3$;
\begin{align}
    \ddot g &=2\frac{(2\beta-3)\zeta(\beta)^2\zeta(\alpha-2)^{\beta-1}}{\pi\cos^2(\beta\pi/2)\Gamma(1-\beta)^2\zeta(\alpha)^{\beta-1}}\frac{\dot g^3}{(2-g)^\beta}-(\beta-1)\frac{\dot g^2}{2-g},\\
    g(0) &=2-\frac{\zeta(\alpha-2)}{\zeta(\alpha)}k^2_0,
\end{align}
for $\alpha>3$ and $\beta<2$;
\begin{align}
    \ddot g &=\frac{4-\alpha}{\alpha-1} \frac{\zeta^2(\beta)}{\pi\zeta^2(\beta-1)\left(-\frac{\zeta(\alpha)}{2\Gamma(1-\alpha)\sin(\alpha\pi/2)}\right)^{\frac{2}{\alpha-1}}}
    \frac{\dot g^3}{(2-g)^{\frac{\alpha+1}{\alpha-1}}}-\frac{2}{\alpha-1}\frac{\dot g^2}{2-g},\\
    g(0) & =2+2\sin(\alpha\pi/2)\frac{\Gamma(1-\alpha)}{\zeta(\alpha)}k^{\alpha-1}_0,
\end{align}
for $\beta>2$ and $\alpha<3$;
\begin{align}
    \ddot g&=\frac{2\beta-\alpha}{\alpha-1}
    \frac{\zeta(\beta)^2}{\pi\cos^2(\beta\pi/2)\Gamma(1-\beta)^2}\left(-\frac{\zeta(\alpha)}{2\Gamma(1-\alpha)\sin(\alpha\pi/2)}\right)^{\frac{2(1-\beta)}{\alpha-1}}
    \frac{\dot g^3}{(2-g)^{\frac{2\beta+\alpha-3}{\alpha-1}}}-2\frac{\beta-1}{\alpha-1}\frac{\dot g^2}{2-g},
    \\
    g(0) & =2+2\sin(\alpha\pi/2)\frac{\Gamma(1-\alpha)}{\zeta(\alpha)}k^{\alpha-1}_0,
\end{align}
\end{widetext}
for $\beta<2$ and $\alpha<3$.

Note that the differential equation for the TFIM, Eq.~\eqref{eq:LZ_opt_diffeq}, shares the same form, 
\begin{equation}
    \ddot g=\frac{1}{2\pi}\frac{\dot g^3g}{(1-g^2)^2}-\frac{2g\dot g^2}{1-g^2}\,.
\end{equation}
To leading order in $1-g$, one can approximate $1+g\approx 2$, leading to
\begin{equation}
    \ddot g=\frac{1}{8\pi}\frac{\dot g^3}{(1-g)^2}-\frac{\dot g^2}{1-g}.
\end{equation}
The same result is obtained when the differential equation is considered in the limits of $\alpha,\beta \to\infty$. Exploiting the simple limits $\zeta(\alpha-2)/\zeta(\alpha)\rightarrow1$ and $\zeta(\beta)/\zeta(\beta-1)\rightarrow1$, it follows that 
\begin{equation}
    \ddot g=\frac{1}{4\pi}\frac{\dot g^3}{(2-g)^2}-\frac{\dot g^2}{2-g}\,,
\end{equation}
where the different constant factors originate from the different position of the critical point, $g_c=2$.

\bibliography{references}

\end{document}